# The Milky Way in Molecular Clouds: A New Complete CO Survey

T. M. Dame[1], Dap Hartmann[2], and P. Thaddeus[3]

*Harvard-Smithsonian Center for Astrophysics,*
*60 Garden Street, Cambridge, MA 02138*

[1] tdame@cfa.harvard.edu

[2] dap@strw.leidenuniv.nl
Presently at Leiden Observatory, P.O. Box 9513, 2300 RA Leiden, The Netherlands

[3] pthaddeus@cfa.harvard.edu

# ABSTRACT

New large-scale CO surveys of the first and second Galactic quadrants and the nearby molecular cloud complexes in Orion and Taurus, obtained with the CfA 1.2 m telescope, have been combined with 31 other surveys obtained over the past two decades with that instrument and a similar telescope on Cerro Tololo in Chile, to produce a new composite CO survey of the entire Milky Way. The survey consists of 488,000 spectra that Nyquist or beamwidth $(1/8°)$ sample the entire Galactic plane over a strip $4°$-$10°$ wide in latitude, and beamwidth or $1/4°$ sample nearly all large local clouds at higher latitudes. Compared with the previous composite CO survey of Dame et al. (1987), the new survey has 16 times more spectra, up to 3.4 times higher angular resolution, and up to 10 times higher sensitivity per unit solid angle. Each of the component surveys was integrated individually using clipping or moment masking to produce composite spatial and longitude-velocity maps of the Galaxy that display nearly all of the statistically significant emission in each survey but little noise.

The composite maps provide detailed information on individual molecular clouds, suggest relationships between clouds and regions widely separated on the sky, and clearly display the main structural features of the molecular Galaxy. In addition, since the gas, dust, and Population I objects associated with molecular clouds contribute to the Galactic emission in every major wavelength band, the precise kinematic information provided by the present survey will form the foundation for many large-scale Galactic studies.

A map of molecular column density predicted from complete and unbiased far-infrared and 21 cm surveys of the Galaxy was used both to determine the completeness of the present survey and to extrapolate it to the entire sky at $|b| < 32°$. The close agreement of the observed and predicted maps implies that only ~2% of the total CO emission at $|b| < 32°$ lies outside our current sampling, mainly in the regions of Chamaeleon and the Gum Nebula. Taking into account this small amount of unobserved emission, the mean molecular column density decreases from ~3 x $10^{20}$ cm$^{-2}$ at $|b| = 5°$ to ~0.1 x $10^{20}$ cm$^{-2}$ at $|b| = 30°$; this drop is ~6 times steeper than would be expected from a plane parallel layer, but is consistent with recent measurements of the mean molecular column density at higher latitudes.

The ratio of the predicted molecular column density map to the observed CO intensity map provides a calibration of the CO-to-$H_2$ mass conversion factor $X \equiv N_{H_2}/W_{CO}$. Out of the Galactic plane ($|b| > 5°$), $X$ shows little systematic variation with latitude from a mean value of $1.8 \pm 0.3$ x $10^{20}$ cm$^{-2}$ K$^{-1}$ km$^{-1}$ s. Given the large sky area and large quantity of CO data analyzed, we conclude that this is the most reliable measurement to date of the mean $X$ value in the solar neighborhood.

*Subject headings:* Galaxy: structure — ISM: clouds — ISM: molecules — radio lines: ISM — solar neighborhood — stars: formation

# 1. Introduction

A large fraction of the interstellar gas in a spiral galaxy such as ours is molecular hydrogen, and much of that is contained in the giant molecular clouds (GMCs), objects with masses of $10^4$-$10^6$ $M_\odot$ and sizes of 50-200 pc at the top of the cloud mass spectrum. The simple, stable diatomic molecule carbon monoxide has played an essential role in the study of GMCs and molecular gas in space generally, because $H_2$ itself, devoid of a permanent electric dipole moment, is almost impossible to observe directly in the cold, generally obscured interstellar regions where molecules form and survive. The lower frequency rotational transitions of CO, in contrast, are readily observed even in quite tenuous molecular gas, and the lowest of these, the 1–0 line at 115 GHz, has become to the radio astronomer the closest molecular analog to the 21-cm line of atomic hydrogen for the study of the interstellar medium. No molecular cloud is free of CO emission, which is generally the most easily observed molecular line, and GMCs throughout the Milky Way can be detected in this line even with a small telescope. It has been shown by three independent methods that the velocity-integrated intensity of the 1–0 line is not only a good qualitative tracer of molecular gas, but a fairly good quantitative tracer as well, providing the column density of $H_2$ to a factor of two or better when averaged over a suitably large region (cf. § 4, and reviews by Solomon & Barrett 1991 and Dame 1993).

Carbon monoxide surveys play a crucial role in many studies of star formation and galactic structure. In conjunction with radio continuum, infrared, and optical observations of H II regions, OB associations, and other Population I objects, they have demonstrated that virtually all star formation occurs in molecular clouds, and high resolution CO observations of dense cloud cores and molecular outflows have contributed greatly to our understanding of how stars form. Since GMCs preferentially form in the arms of spiral galaxies (e.g., Loinard et al. 1999; Aalto et al. 1999) and it is possible to resolve the kinematic distance ambiguity for many of those in the inner Milky Way by appeal to the associated Population I, CO surveys have helped to refine our knowledge of the spiral structure of our system (e.g., Dame et al. 1986, Grabelsky et al. 1988). By virtue of the precise kinematic information they provide, CO surveys have also been of great value in the interpretation of Galactic continuum surveys from satellite observatories such as GRO, ROSAT, and COBE.

Because molecular clouds are so large, contain such a large fraction of the interstellar gas and dust, and are the source of so many conspicuous young objects, they can be detected across the electromagnetic spectrum. GMCs are the source of much of the diffuse Galactic gamma ray emission (Hunter et al. 1997), and two of the rare gamma ray repeaters have recently been found to be associated with very massive GMCs (Corbel et al. 1997, 1999). GMCs can be detected in absorption against the diffuse X-ray background (Park, Finley, & Dame 1998), and associated supernova remnants are often strong X-ray emitters (e.g., Seward et al. 1995; Slane et al. 1999). If close enough, even quite modest molecular clouds are conspicuous as classical dark nebulae (§ 3.2), and distant, visually obscured GMCs can be detected as dark nebulae in the near infrared against the Galactic bulge (Kent, Dame, & Fazio 1991). Warm dust in molecular clouds is responsible for much of the diffuse far infrared emission observed by IRAS (Sodroski et al. 1997), and associated Population I objects are responsible for many of the strong IRAS point sources, masers, and radio continuum sources.

The work described here is the culmination of Galactic CO surveys done over two decades with two small millimeter-wave telescopes designed to undertake the first uniform Galactic survey of molecular clouds, one originally at Columbia University in New York City, now in Cambridge, Massachusetts, the other at the Cerro Tololo Interamerican Observatory in Chile. The angular resolution of these 1.2 m telescopes, ~8.5' at 115 GHz, yields a linear resolution of ~20 pc at the Galactic center, large enough so that major segments of the Galactic plane can be covered in a reasonable amount of time, yet adequate to resolve the larger GMCs throughout the Milky Way.

After preliminary surveys of the Galactic plane (Burton et al. 1975, Scoville & Solomon 1975, Cohen & Thaddeus 1977) and large local clouds such as those in Orion (Kutner et al. 1977) revealed the vast extent of CO emission on the sky, it became clear that even with telescopes as small as ours a sensitive, well-sampled survey of the entire Galaxy would require many years. For that reason, during the period 1979-1986, the telescopes carried out a series of "superbeam" surveys in which angular resolution was sacrificed for the sake of coverage and speed. Most of these studies were conducted at an effective angular resolution of 0.5°, achieved by stepping through a 4 x 4 grid of positions on the sky separated by $1/8°$, slightly less than one beamwidth, during the accumulation of a single spectrum. These low-resolution surveys, in total comprising over 31,000 spectra and sampling nearly a fifth of the entire sky (~7700 deg$^2$), were combined into the first complete CO map of the Milky Way (Dame et al. 1987).

Even before these low-resolution surveys were undertaken, the two telescopes began to survey the Galaxy and its nearest neighbors at several times higher angular resolution—typically every beamwidth, but sometimes half or twice that—and at 5-10 times higher sensitivity per solid angle. Such observations now cover the entire Galactic plane over a 4°-10° band in latitude, all large local clouds at higher latitude, as well as the Large Magellanic Cloud and M 31. In this paper, we combine all of the full-resolution observations, a total of 488,000 spectra, into a new composite CO map of the Galaxy. The original 0.5° map is still used in a few higher-latitude regions where full-resolution observations have not yet been made. Some of the full-resolution data presented here have already been published as separate surveys of particular clouds or regions (Table 1); the rest are recent studies of the first and second Galactic quadrants, the Taurus and Orion clouds, and the λ Orionis ring.

Although the present survey covers only 45% of the 20,626 deg$^2$ of the sky within 30° of the Galactic equator, we will show that it is probably very nearly complete for clouds larger than a few degrees on the sky. Using complete and unbiased far infrared surveys as tracers of total gas column density and the new Leiden-Dwingeloo 21 cm survey (Hartmann & Burton 1997) as an inventory of the atomic gas, we will construct a map of inferred molecular column density that agrees very well with our observed CO map. Within the past few years, this type of analysis has led to the discovery of a number of new clouds away from the Galactic plane that are included in the composite map here.

We begin the next section (§ 2) with a discussion of the two telescopes, including a brief history, a description of their current instrumentation, and a discussion of the data acquisition and reduction employed in the various surveys. We describe how differences among the surveys were reconciled in constructing the composite maps; special attention is paid to slight differences in the calibration of the various surveys and between telescopes, since correction for this variation is crucial to achieve a common intensity scale. In § 3 two standard exhibits of the composite survey are then presented: (*i*) a spatial map integrated over all velocity, and (*ii*) a longitude-velocity map integrated over latitude. These maps will be discussed in detail, with special focus on the new surveys mentioned above. By appealing to existing H I and far-infrared all-sky surveys, we show in § 4 that, in spite of gaps in our sky coverage, little CO emission at $|b| < 32°$ is likely to have been missed by the composite survey here. Finally, in § 5 we investigate both the CO-to-H$_2$ conversion factor and the mean molecular column density as functions of Galactic latitude.

## 2. Observations and Analysis

### 2.1 Observations

Most of the data presented here were obtained with our two telescopes in their current configurations, as described in Dame et al. (1993) for the northern instrument and in Cohen (1983) for the southern. Because some of the surveys were obtained as long ago as 1980, however, we review below the major changes of instrumentation that have occurred since that time.

Both antennas are diffraction-limited Cassegrain systems with 1.2 m diameter parabolic primaries and 17.8 cm hyperbolic secondaries. Prior to 1983, the northern telescope had a smaller 15.2 cm secondary which yielded a smaller beam but also a lower beam efficiency. In early 1994, the three 1" diameter tubular support arms of the secondary were replaced by solid rectangular arms 1"x 3/8" in cross section, to achieve lower sidelobes and a slightly higher beam efficiency. The purpose of this refinement was to facilitate the study of very weak CO emission high above the plane in the inner

Galaxy from the thick component of the molecular disk (Dame & Thaddeus 1994), for which sidelobe contamination from the much more intense, extended emission near the plane is a potential problem. The northern telescope in its current configuration has a beamwidth of $8'.4 \pm 0'.2$ (FWHM; Appendix A). The southern telescope has nearly the same beamwidth to within the uncertainties: $8'.8 \pm 0'.2$ (Bronfman 1986).

Spectral line intensities are calibrated by measuring the receiver response to two thermal sources of known intensity: a standard blackbody Eccosorb chopper wheel at ambient temperature, and the sky, whose radiation temperature at the zenith is about 90 K on a cold, dry winter day. The temperature of the chopper wheel, which is rotated just in front of the feed horn at 20 Hz for 1 s before each observation, is controlled and monitored by the telescope computer, and the effective blackbody temperature of the sky is determined from the two-layer atmospheric model of Kutner (1978). The parameters of the model—the temperature and opacity of atmospheric water vapor and the fraction of forward power on the sky—are determined by antenna tipping every 6 hr when the atmosphere is clear and steady, and more often when the weather is changing. This standard chopper-wheel calibration (Penzias & Burrus 1973) yields units of $T_A^*$: antenna temperature corrected for atmospheric attenuation, ohmic losses, and rearward spillover. Intensities reported here are further divided by the main beam efficiency (the fraction of forward-projected power in the main beam, out to 11') to obtain the main beam brightness temperature, $T_{mb}$ (Downes 1989). In the Raleigh-Jeans limit, $T_{mb}$ measured for a uniform source that just fills the main beam equals its effective blackbody temperature. As discussed in Appendix A, both the northern and southern telescopes have main beam efficiencies of $0.82 \pm 0.04$.

During the nearly two decades over which the present data were obtained, measured $T_A^*$ on the 1.2 m telescopes for both point and extended sources have changed slightly due to changes in feed horn, receiver, focus, secondary supports, and assumed oxygen opacity. However, since these changes were accurately tracked with test sources and overlapping surveys, reported $T_{mb}$ values (often called $T_R^*$ or "$T_R$ for a source that just fills the main beam" in past papers) have only changed once, by ~20%, owing to a reevaluation of the beam efficiencies by Bronfman et al. (1988). Because the calibration of the northern telescope in its present configuration is so well understood, we compared its measured intensities to those of previous surveys with both telescopes in order to put all surveys on a common and self-consistent scale of $T_{mb}$.

Most of the data from the northern telescope were obtained with an extremely sensitive SIS heterodyne receiver which was installed in 1983 (Pan 1984). Its single-sideband noise temperature of ~65 K yields total system temperatures referred to above the atmosphere of 400-800 K for the normal range of elevations observed, 30°-75°. Some early data on the northern telescope were obtained with an uncooled Schottky diode mixer (Cong, Kerr, & Mattauch 1979) with a noise temperature of 900 K. All of the southern data were obtained with a liquid-nitrogen cooled Schottky diode receiver (385 K, SSB).

All survey data were obtained with one of three 256-channel filterbank spectrometers of NRAO design (Mauzy 1974), identical except for their filter widths. The most heavily used filterbank has 500 kHz filters, which at 115 GHz yield a velocity resolution of 1.3 km s$^{-1}$ and a total velocity range of 332 km s$^{-1}$—wide enough to cover the full range of Galactic CO velocities with adequate baseline except toward the Galactic center, where each position required two velocity settings. The 500 kHz filterbank was used for the surveys of the southern Galactic plane (and the LMC), then transferred to Cambridge in 1988 where it was used for the M31 survey (Dame et al. 1993) and most subsequent work with the northern telescope. The northern telescope is also equipped with a 250 kHz filterbank which was used for all surveys prior to 1988 and since then has been run in parallel with the 500 kHz filterbank, providing higher velocity resolution (0.65 km s$^{-1}$) and redundancy in the central 166 km s$^{-1}$ of the observed band. The southern telescope has always been equipped with a 100 kHz filterbank; the high velocity resolution (0.26 km s$^{-1}$) but relatively narrow coverage (66 km s$^{-1}$) of this spectrometer has mainly been dedicated to the study of local clouds well out of the Galactic plane, such as those in Ophiuchus.

To obtain flat spectral baselines close to the Galactic plane where emission typically covers a large range in velocity, spectra were acquired by position switching every 15 s between the source position (ON) and two emission-free reference positions (OFFs) selected by the telescope control program to straddle the ON in elevation. The fraction of the time spent on each OFF was adjusted so that the time-weighted average system temperature at the OFFs was equal to that at the ON, resulting in baselines that were very flat, and residual offsets that were typically less than 1 K. This offset was generally removed by fitting a straight line to the emission-free ends of the spectrum. The OFF positions used for our surveys are discussed in Appendix B.

Away from the plane, in those regions where only one or two relatively narrow CO lines are found, frequency-switching by 10-20 MHz at a rate of 1 Hz was often used instead of position switching. Since spectral lines remain within the range of the spectrometer in both phases of the switching cycle, data could be obtained twice as fast as with position switching, although higher order polynomials, typically 4th or 5th order, were required to remove the residual baseline. A telluric emission line from CO in the mesosphere, variable in both intensity and LSR velocity, is detected in frequency-switched spectra; because the LSR velocity of the line could be predicted exactly, blending with Galactic emission could be avoided by appropriate scheduling of the observations. In a few cases, most notably the recent large-scale surveys of the λ Orionis ring (Lang et al. 2000) and the Taurus clouds (S. T. Megeath, T. M. Dame, & P. Thaddeus 2001, in prep.), a model of the telluric line was fit daily to spectra free of Galactic emission and used to remove the line from all spectra.

Observational parameters such as *rms* noise, sampling interval, and velocity resolution were held constant within each of the 37 surveys listed in Table 1; the sky coverage of each is shown in Figure 1. A uniform *rms* noise within each survey was achieved by automatically adjusting the integration time of each scan depending on the instantaneous system temperature as determined from the 1 s calibration. Integration times varied from 20 s to 5 min, depending on each survey's target *rms*, weather conditions, receiver noise, and source elevation. The *rms* noise levels of individual surveys, listed in column 9 of Table 1, vary from 0.12 to 0.43 K. Comparing the surveys based on these *rms* values alone can be misleading, given that differences in spatial and velocity sampling exist as well. A better comparison is with the *rms* noise per unit solid angle and velocity interval, given in the next column of Table 1.

## 2.2 Synthesis of the Data

A composite CO survey was constructed from the 37 individual surveys listed in Table 1. The individual surveys contain a total of 487,890 spectra, 80% of which are on a grid sampled slightly better than every beamwidth ($1/8°$) or every half-beamwidth ($1/16°$). The total area covered by the composite survey is 9,353 $deg^2$—more than one-fifth of the entire sky and nearly one half of the area within 30° of the Galactic plane.

Since the individual surveys are stored as 3-dimensional (*v-l-b*) datasets in FITS format (Wells et al. 1981), it is fairly straightforward to achieve by interpolation a composite dataset on a uniform spatial and velocity grid. Various versions of this dataset, as well as subsections of it and the individual surveys, are available for distribution. Here we present two of its most useful projections: (*i*) a spatial map integrated over velocity (hereafter, the $W_{CO}$ map), and (*ii*) a longitude-velocity map integrated over latitude (the *l,v* map). The most direct way to produce these maps is to integrate the composite data cube, but doing so results in maps with non-uniform noise; it is difficult with these to display the weaker features detected by the more sensitive surveys without displaying excessive noise elsewhere in the map. Therefore, rather than simply integrating the composite data cube, we produced the composite maps by integrating each survey individually, using one or more of the following three techniques where necessary to suppress noise:

(i) The velocity integration range was always chosen to cover less than the full range of the spectrometer; it was generally restricted to a range ~30 km s$^{-1}$ wide about zero at $|b| >$ 3°, and to a wider range permitted by Galactic rotation at lower latitudes.

(ii) For all of the Galactic plane surveys and some others with complicated velocity structure, a heavily-smoothed and therefore low-noise version of each survey was used to restrict the velocity-integration windows even further. Each survey was smoothed to a resolution of 0.3° x 3 km s$^{-1}$ (or 0.6° x 3 km s$^{-1}$ for the surveys with 0.25° sampling), and integration of the original survey was then taken only over channels for which the intensity in the smoothed survey exceeded 3 times its *rms* noise (typically ~0.05 K). This "masked moment" method was originally developed to analyze 21 cm surveys of galaxies with limited signal-to-noise, and has recently been applied to extragalactic and Galactic CO surveys (e.g., Adler et al. 1992; Digel et al. 1996; Loinard et al. 1999).

(iii) The most drastic noise reduction was applied to the velocity-integrated superbeam maps, which were only used in some high-latitude regions of the composite $W_{CO}$ map (gray regions in Figure 1). These maps needed to be essentially free of noise spikes, since a single such spike—one half degree across—would appear highly significant in the composite map, which is largely beamwidth sampled. The individual superbeam surveys (listed in Table 1 of Dame et al. 1987) were clipped at 3 times their *rms* noise levels (i.e., channels with absolute intensities less than 3-σ were set to zero before integrating over velocity).

Many of the weaker features in the integrated maps were checked to assure that they corresponded to identifiable spectral lines in the corresponding spectra, and not to baseline fluctuations or statistical noise.

The individual integrated maps were subsequently combined, the better-quality survey being used in regions where several overlapped. The composite $W_{CO}$ map was constructed with a spacing of 0.125° in both *l* and *b*—slightly better than one beamwidth. Surveys with half-beamwidth spacing were smoothed by convolution with a Gaussian with FWHM of 0.125°; those with coarser sampling were interpolated to the 0.125° grid.

The composite *l,v* map required only the large surveys of the four Galactic quadrants (numbers 8, 17, 31, and 36 in Table 1) plus those of the Galactic center and Carina (2 and 33, respectively). Except for the third quadrant survey, emission was integrated using moment masking over a 4° strip of Galactic latitude centered on the plane. As Figure 1 shows, the centroid and width of the latitude coverage of the third quadrant survey varied with longitude; this survey was integrated over its full latitude extent, which was typically ~3.5° centered near *b* = -1°. The latitude integration range for all surveys was sufficient to include essentially all emission beyond the local spiral arm (i.e., at $|v| > 20$ km s$^{-1}$). The composite map was constructed on a 0.125° x 1 km s$^{-1}$ grid, and was subsequently smoothed in velocity to a resolution of 2 km s$^{-1}$ (slightly more than was necessary to resolve filterbank differences) and in longitude to a resolution of 0.2° (slightly more than the $W_{CO}$ map). This extra smoothing enhanced weak extended features such as the outer spiral arm at negative velocities in the first quadrant.

## 3. The Composite Maps

The composite spatial and longitude-velocity maps of the Galaxy are shown in Figures 2 and 3, respectively. As described above, the clipping or moment masking parameters for each component survey were carefully adjusted so that nearly all statistically-significant emission in each survey appears in the composite maps, but little noise. The total dynamic range of these maps is approximately a factor of 100, with each distinct color change corresponding to about a factor of two change in intensity, as indicated by the color bars. The weakest features displayed in the maps are a factor of 3-4 below the lowest contours of the corresponding superbeam maps of Dame et al. (1987); with a factor of 2-4 better angular resolution, the new maps provide an order of magnitude gain in sensitivity per unit solid angle over the previous superbeam maps.

### 3.1 Individual Regions

Figures 2 and 3 exhibit with great clarity the large-scale distribution and kinematics of the molecular clouds in the Milky Way. Moving from left to right in these maps, we will discuss the more obvious clouds and regions, describe any special aspects of our observations in each region, and note published, large-scale CO surveys carried out with other telescopes. Most of the regions and objects discussed are identified in the finder charts of Figures 2 and 3.

*(i) Taurus Dark Clouds*

A superbeam survey of the well-known system of dark clouds in Perseus, Taurus, and Aurgia, lying below the plane near the Galactic anticenter, was conducted by Ungerechts and Thaddeus (1987). They identified three main clouds in this extensive nearby complex. The large, roughly parallel elongated clouds running above and below the Per OB2 association lie at about the same distance as that association (~350 pc), and presumably are associated with it; the upper cloud contains the California Nebula (NGC 1499) and NGC 1579, the lower one IC 348 and NGC 1333. A third cloud, forming the dark nebulae in Taurus and Auriga, is closer (~140 pc) and more quiescent; it has proven to be one of the best regions for studying low-mass star formation. A large section of this cloud (~40 deg$^2$) was mapped in $^{13}$CO with the Nagoya 4 m telescopes by Mizuno et al. (1995). Our beamwidth-sampled survey of the entire complex was completed during the 1998-99 observing season, and will be described in a forthcoming paper (S. T. Megeath, T. M. Dame, & P. Thaddeus 2001, in prep.). This new survey is second only to the full 2nd quadrant survey in terms of total area observed and total number of spectra (Table 1).

*(ii)   Second Quadrant*

The largest individual CO survey, covering the entire second quadrant as well as small parts of the first and third quadrants, required 6-8 hours per day on the northern telescope during the past 9 observing seasons (1992-2000). As the *l,v* map (Fig. 3) shows, the emission in this region is tightly confined to two parallel lanes in velocity, one corresponding to local material near 0 km s$^{-1}$ and the other to the Perseus spiral arm at velocities near -50 km s$^{-1}$. Owing to its proximity and the lack of kinematic distance ambiguity beyond the solar circle, the Perseus arm in the second quadrant is the best place in the Galaxy for studying a large population of molecular clouds at roughly the same distance. All of the emission associated with the Perseus arm was sampled every half beamwidth (survey 17; Fig. 1), while the very extensive local emission was mapped every other beamwidth (survey 18). A preliminary section of the Perseus arm survey covering the W3-4-5 star-forming complex was described by Digel et al. (1996). The most massive complex in the Perseus arm, that associated with the SNR Cas A and NGC 7538, was mapped even earlier with somewhat different observing parameters by Ungerechts, Umbanhowar, & Thaddeus (2000; survey 14). A large section of the second quadrant in Cygnus, Cepheus, and Cassiopeia (*l* = 80° to 130° over a strip 18°-30° wide in latitude) was mapped in $^{13}$CO with the Nagoya 4 m telescopes (Dobashi et al. 1994; Yonekura et al. 1997), but these surveys cover in velocity only the local emission.

The second quadrant was also the subject of the largest CO survey so far conducted with the FCRAO 14 m telescope. Heyer et al. (1998) used the 15-element QUARRY array receiver to survey an 8°-wide strip of latitude centered on the Perseus Arm between *l* = 102° and 141°. Since a large fraction of their survey overlaps our own half-beamwidth survey, it is of interest to compare the two. Such a comparison is shown in Figure 4: the FCRAO W$_{CO}$ map in (*a*), our corresponding map in (*b*), and the FCRAO map smoothed to our angular resolution in (*c*). The striking similarity of even very small features in the CfA map (Fig. 4*b*) with those in the smoothed FCRAO map (Fig. 4*c*) strongly supports our conclusion that the composite W$_{CO}$ map displays almost all significant emission but few features attributable to noise; more generally, it provides a very stringent check on the integrity of both surveys. The general similarity of the CfA map with even the unsmoothed FCRAO map in Figures 4*a*, which has 10 times higher angular resolution, is a demonstration of the importance of sensitivity per unit solid angle (column 10 in Table 1): the sensitivity per unit solid angle of the FCRAO survey is about twice that of the CfA survey, but in the CfA survey this difference is removed by the moment masking applied to Figure 4*b* (see § 2.2).

*(iii)   Cepheus & Polaris Flares*

The Cepheus and Polaris Flares, stretching above the Galactic plane toward the north celestial pole, form a system of high-latitude clouds so extensive that they were noticed as a region of low galaxy counts by Hubble (1934). The Cepheus Flare was first mapped at superbeam resolution by Grenier et al. (1989), who noted that this cloud and another large filamentary cloud closer to the plane in Cassiopeia appeared to form a loose ring ~12° in diameter surrounding a region of soft X-rays and radio continuum emission, possibly a supernova remnant. These Flares point toward an even higher-latitude and possibly related system of clouds in Ursa Major, lying partially beyond the latitude limit of our map near $l \sim 145°$. The technique of using far-infrared emission as a total gas tracer in order to calibrate the $N_{H_2}/W_{CO}$ ratio, which we apply to the whole Galaxy in the next section, was first suggested and applied locally by de Vries, Heithausen, & Thaddeus (1987) to the Ursa Major clouds. Heithausen et al. (1993) combined the Ursa Major observations with a beamwidth-sampled map of the Polaris Flare (Heithausen & Thaddeus 1990) and their own observations of the nearby Camelopardalis complex in order to study the infrared properties and overall mass surface density of molecular cirrus clouds.

*(iv) IR-Predicted Clouds*

The surveys far below the plane in the second quadrant (No. 11, 12, & 13) as well as the other 7 surveys colored purple in Figure 1 (Nos. 5, 6, 23, 25, 28, 29, and 30) were carried out within the past few years, following the prediction described in § 4 of molecular clouds in those regions. As we will show, in all but one case (survey 23) molecular clouds were detected in fairly good agreement with those predicted. Only the Aquila South region (survey 6) had been observed previously with the northern telescope, at a reduced resolution of 0.5° (Lebrun 1986). Comparison of the Aquila South region in the superbeam map of Dame et al. (1987; their Fig. 2) with the same region in Figure 2 provides a good demonstration of the improved angular resolution and sensitivity here: the large complex of relatively dim clouds seen in this region of Figure 2 lies completely below the lowest contour of the superbeam map.

*(v) First Quadrant*

The intense emission from the inner spiral arms of the Galaxy in the first quadrant has been the subject of numerous CO surveys over the past two decades. Preliminary observations by Cohen & Thaddeus (1977) were expanded to a beamwidth-sampled survey of the plane by Cohen, Dame, and Thaddeus (1986); roughly the same area has recently been mapped in the CO (J=2–1) line with the same angular resolution and grid spacing by Sakamoto et al. (1997). The first large-scale superbeam survey, with 1° angular resolution, was conducted in the first quadrant by Dame & Thaddeus (1985), and a subsequent survey at 0.5° resolution was included in the composite survey of Dame et al. (1987). The very sensitive observations included in the present composite map were begun in 1991 as a series of narrow, Nyquist-sampled strips perpendicular to the Galactic plane (survey 7); Dame & Thaddeus (1994) used this survey—with the highest sensitivity per unit solid angle of any in the composite map (see column 10 of Table 1)—to detect a faint, extended molecular disk in the inner Galaxy about three times thicker than that of the dense central CO layer, and comparable in thickness to the central H I layer. Subsequently, observations with the same sensitivity but beamwidth sampling were extended to cover most of the first quadrant over ~10° in Galactic latitude (survey 8). In addition to the thick molecular disk, other features revealed by the new observations include a faint substratum of emission in the gas hole within ~3 kpc of the Galactic center (at $v \geq 50$ km s$^{-1}$ and $l \leq 20°$ in Figure 3) and a distant spiral arm lying beyond the solar circle and spanning most of the first quadrant (at $v \leq$ -20 km s$^{-1}$ in Figure 3).

*(vi) Ophiuchus, Lupus, & R CrA*

The Ophiuchus and Lupus clouds lie well above the plane near the center of the map. Ophiuchus contains one of the nearest, best studied regions of star formation; Lupus, like Taurus, appears to be forming only stars of low to intermediate mass, including more than 100 T Tauri stars recently discovered by Krautter et al. (1997). Both clouds lie at similar distances (~150 pc) and have similar masses (~$10^4$ M$_\odot$); de Geus (1992) has shown that both lie near the edge of an expanding H I shell surrounding the Upper Scorpius subgroup of the Sco-Cen OB association. Lupus is the largest local cloud in our map that has only been observed at 0.5° resolution (Murphy, Cohen, & May 1986); roughly 25 deg$^2$ in each cloud has recently been mapped in $^{13}$CO with the Nagoya 4 m telescopes (Ophiuchus: Nozawa et al. 1991; Lupus: Tachihara et al. 1996).

Across the Galactic plane from Ophiuchus and Lupus, and lying at about the same distance (130 pc; Marraco & Rydgren 1981), the small, isolated R CrA cloud is named after the bright emission-line star at its core. In addition to this A star and the B9 star TY CrA, near-infrared surveys have shown that lower mass stars have formed throughout the cloud (e.g. Wilking et al. 1997); an obscured cluster of such stars around R CrA was discovered by Taylor & Storey (1984) and dubbed the Coronet. In addition to our observations with 0.25° sampling, the dense core has been mapped in CO at 2.6' resolution by Loren, Peters, and Vanden Bout (1974) and in C$^{18}$O at 40" resolution by Harju et al. (1993).

*(vii) Galactic Center*

The Galactic center region is distinguished in the W$_{CO}$ map (Fig. 2) by its very high intensity, roughly 4 times that seen elsewhere along the plane, and in the *l,v* map (Fig. 3) by its well-known large non-circular motions, extending to ±250 km s$^{-1}$. Among the most striking features in this region of the *l,v* map are the expanding 3 kpc arm, an inclined linear feature with a velocity of -55 km s$^{-1}$ toward $l = 0°$, and the pair of intense vertical features at $l = 3°$ and $5°$, each extending from $v = 0$ km s$^{-1}$ to beyond 150 km s$^{-1}$. On the plane of the sky these objects appear as fairly well-defined GMCs, with radii of

~75 pc at the distance of the Galactic center, but their linewidths are ~10 times that of typical GMCs elsewhere in the Galaxy.  The nature of these so-called wide line clouds has been discussed by Stacy et al. (1988) and Kumar & Riffert (1997); the one at *l*=3°, also known as Bania Clump 2, was discussed by Stark & Bania (1986).  The southern telescope has mapped the Galactic center in both CO (Bitran et al. 1997) and $C^{18}O$ (Dahmen et al. 1997); other large-scale CO, $^{13}CO$, $C^{18}O$, and CO (J=2–1) surveys of the Galactic center region are listed in Table 1 of Bitran et al. (1997).

*(viii)   Chamaeleon  &  Carina*

The Chamaeleon dark clouds lie well below the plane in the fourth quadrant.  A recent study of the distribution of reddening with distance toward the three main clouds of this complex derived a common distance of ~160 pc (Whittet et al. 1997), quite similar to the Lupus, Ophiuchus, and R CrA clouds.  As we will show in the next section, the far-infrared and H I emission in this region suggest that significantly more molecular gas may lie outside our current sampling, which is rather tightly confined to the main clouds (see Fig. 2).  Molecular gas in this region unobserved by us is also suggested by (*i*) the "sheet-like" distribution of reddening material detected by Whittet et al. (1997), (*ii*) the widely-distributed population of pre-main sequence T Tauri stars recently detected by ROSAT (Alcalá et la. 1995; Covino et al. 1997), and (*iii*) a corresponding population of small dense molecular clouds revealed by the $^{13}CO$ survey of Mizuno et al. (1998).

The Carina nebula lies near the tangent point at longitude 280° of the Carina spiral arm, one of the best-defined structural features of the molecular Milky Way.  The loop structure characteristic of a spiral arm is unmistakable in the Carina region of the *l,v* map (Fig. 3).  Owing to the absence of ambiguity in kinematic distances beyond the solar circle and the relatively low overall density of clouds there, the far side of the arm at positive velocity can be traced to very large distances.  The most distant clouds in this arm, near l~325°, v~30 km s$^{-1}$, lie more than 17 kpc from the Sun.

*(ix)  Third  Quadrant*

The distribution of molecular gas beyond the solar circle in the third Galactic quadrant is distinctly different from that in the second quadrant. The extensive system of filamentary local clouds seen in the second quadrant is all but absent; there are instead a number of quite massive and well-defined complexes lying 1-2 kpc from the Sun, including those associated with CMa OB1 and the Rosette Nebula (Blitz 1978), Mon OB1 (Oliver, Masheder, & Thaddeus 1996) and Gem OB1 (Stacy & Thaddeus 1988; Carpenter, Snell, & Schloerb 1995).  In addition, the massive, unusually cold cloud studied by Maddalena & Thaddeus (1985) lies just below the plane between CMa OB1 and the Rosette; Lee, Snell, & Dickman (1994) recently mapped this cloud at 10 times higher angular resolution.  A third-quadrant extension of the Perseus Arm is visible in the *l,v* map (Fig. 3), but it is traced by fewer and generally fainter clouds than in the second quadrant, where the arm is closer to us and at a smaller Galactic radius.

*(x) Orion Region*

Three of the most extensively studied molecular clouds, those associated with the Orion nebula, NGC 2024 (Orion B), and Mon R2, are conspicuous well below the plane near $l \sim 210°$. A new CO survey of this region by the northern telescope (Table 1), just completed during the 1999-2000 observing season, improves on the coverage and sensitivity of the previous survey by Maddalena et al. (1986). The entire Orion A cloud has been mapped in $^{13}$CO with both the Bell Laboratories 7 m telescope (Bally et al. 1987) and the Nagoya 4 m telescopes (Nagahama et al. 1998), and both the A and B clouds have been surveyed in the 2–1 transition of the normal isotopic species with the Tokyo-NRO 60 cm telescope (Sakamoto et al. 1994). The Mon R2 cloud has been surveyed in the normal 1–0 CO line with the FCRAO 14 m telescope (Xie & Goldsmith 1994). The Orion clouds in Figure 2 appear connected to the Galactic plane by two remarkably long and thin molecular filaments, and our recent observations (surveys 28 & 29) suggest that these filaments may extend through and above the plane. The two filamentary features lying above the plane near $l \sim 213°$ are inclined to the plane by about the same angle as the Orion filaments, and have the same velocity as the northern one: $\sim 10$ km s$^{-1}$.

Adjacent to the Orion complex toward lower longitude, the well-known ring of gas and dust surrounding the H II region S264 and its ionizing star λ Ori is clearly traced by molecular gas. The lack of CO emission within this ring and a systematic velocity gradient across it suggests that the molecular gas is mainly confined to an expanding torus inclined to the line of sight rather than to a spherical shell (Lang et al. 2000). The W$_{CO}$ map (Fig. 2) hints that a similar ring-like structure may lie nearby in the plane, at $l \sim 180°$, where it could be partially masked by unrelated material along the line of sight. As shown in the finder diagram of Figure 2, the better-defined lower half of this apparent ring surrounds the very old ($\sim 10^5$ yrs) and large ($\sim 3°$ diameter) supernova remnant S 147 (Kundu et al. 1980). This region is included in a large $^{13}$CO survey recently completed toward the Galactic anticenter ($170° < l < 196°$, $|b| < 10°$) with the Nagoya 4 m telescopes (Kawamura et al. 2000).

## 3.2 Large-Scale Structure

The overview of the molecular Galaxy provided by the composite maps allows structural features of the system to be distinguished which are often not apparent in the surveys of the individual regions. It provides a perspective which can suggest relationships between clouds and regions widely separated on the sky, and it can help in deciding whether a particular cloud is unusual in, for example, its structure, mass, internal motions, or Galactic location.

The most prominent feature in Figure 2 is the thin, intense ridge of emission extending ~60° to either side of the Galactic center. The $l,v$ map (Fig. 3) shows that the high intensity of this ridge results from integration over many molecular clouds in the so-called molecular ring (see Fig. 3 finder chart), the broad peak in molecular density roughly halfway between the Sun and the Galactic center. The asymmetric appearance of the molecular ring in the $l,v$ map indicates that it is composed of several inner spiral arms, perhaps rooted in a central bar (e.g., Fux 1999). The rapid falloff in the density of molecular clouds outside the molecular ring is evident by comparing the non-local ($|v|$ > 20 km s$^{-1}$) emission toward the inner Galaxy ($l$=270° to 90°) with that toward the outer Galaxy ($l$=90° to 270°).

Most of the stronger (red-to-white) peaks in the $l,v$ map correspond to giant molecular complexes with masses in the range $10^5 - 10^6$ M$_\odot$. Two of the nearest of these objects are those associated with the supernova remnants W44 in the Sagittarius arm ($l$ = 35°, $v$ = 45 km s$^{-1}$) and Cas A in the Perseus Arm (110°, -55 km s$^{-1}$). Even though each of these can be decomposed into smaller clouds, Figure 3 leaves little doubt that each can be considered a single, well-defined object with a total mass of ~$10^6$ M$_\odot$ and internal motions in excess of 15 km s$^{-1}$. Studies of the molecular cloud mass spectrum, particularly in the Perseus arm (Dame 1983, Casoli et al. 1984; Digel et al. 1996), indicate that a large fraction of the total Galactic CO emission and molecular mass is contained in such large, fairly well-defined molecular complexes.

Clouds within 1 kpc of the Sun appear in the $l,v$ map at $|v|$ < 20 km s$^{-1}$. Such clouds do not appear randomly distributed in velocity, but form a rather well-defined lane in Figure 3, slightly displaced by differential Galactic rotation to positive velocity in the first and third quadrants, and to negative velocity in the second and fourth. Figure 5 shows that most of the large local clouds lying beyond the latitude integration range of Figure 3 also fall within this same low-velocity lane. The well-defined nature of this lane is probably a result of the tight confinement of the nearby molecular gas to the Local spiral arm or spur in which the Sun is located; toward the inner Galaxy, the Local arm emission is well separated in velocity from that of the Sagittarius-Carina arm (e.g., Grabelsky et al. 1988), and owing to the absence of a distance ambiguity in the outer Galaxy, it is even more cleanly separated from the Perseus Arm emission. As has been long suggested (e.g., Bok 1959), the rather similar Cygnus X ($l$ ~ 80°) and Vela ($l$ ~ 270°) regions, lying along the Galactic plane ~180° apart, may be the directions in which we view the Local arm tangentially. Beside being very intense in CO, both regions contain strong concentrations of Population I objects spread over a considerable range of distances.

The kinematics of the local molecular gas may also contain evidence of supershells of the kind seen in H I (e.g., Heiles 1984). In the $l,v$ map (Fig. 3) the local emission over much of the second quadrant ($l \sim 100°$ - $165°$) appears bifurcated in velocity into components near 0 and -12 km s$^{-1}$. Two local velocity components are also seen in 21 cm observations over most of the Galactic plane (Lindblad 1974). The 21 cm component corresponding to the clouds at -12 km s$^{-1}$ was identified with the Local spiral arm by Lindblad (1974); the other component, Lindblad's Feature A, was modeled by Lindblad et al. (1973) and Grape (1975) as an expanding shell of cold gas surrounding the Sun and possibly related to Gould's Belt. Blaauw (1991) proposed that this shell, with an estimated expansion age of 60 million years (Lindblad et al. 1973), was formed by high-mass stars in the old Cas-Tau OB association.

The local molecular clouds can be associated with Gould's Belt not only by their correlation in velocity with Lindblad's Feature A, but also by their spatial distribution. Except for the prominent Cepheus and Polaris Flares, most of the major local molecular clouds appear to follow Gould's Belt: the maximum northern extension of the Belt is marked by the clouds in Lupus, Ophiuchus, and Aquila while its southern extension is marked by those in Orion and Taurus.

The spatial distribution of low-velocity CO emission is compared to an optical panorama of the inner Galaxy in Figure 6. Because some molecular clouds as far away as 2-3 kpc can be seen as dark nebulae (Huang, Dame, & Thaddeus 1983; Dame & Thaddeus 1985), the integration range of the CO map in Figure 6*b* varies with longitude in order to include emission with near kinematic distances less than 2.5 kpc. While this procedure eliminates emission from clouds deep in the inner Galaxy, a thin ridge of emission near $b = 0°$ remains, owing to clouds near the solar circle on the far side of the Galaxy. Except for these clouds, which are too far away to appear as dark nebulae, there is a very tight correlation between the CO emission and optical obscuration: nearly all dark clouds are seen in CO, and vice versa. This correlation argues strongly that CO is faithfully tracing all regions where the gas and dust densities are sufficiently high to allow molecules to form and to produce an appreciable dark cloud. There is little evidence from the optical obscuration for clouds that are deficient in CO, or so cold that CO fails to radiate at 115 GHz.

## 4. Completeness

It is now well established that the far-infrared cirrus emission observed by IRAS (Low et al. 1984) is well correlated with the column density of H I over much of the sky at high Galactic latitude (e.g., Boulanger, Baud, & Van Albada 1985, Désert, Bazell, & Boulanger 1988; see also Fig. 7*c* below). In the small number of high-latitude molecular clouds that have been studied in detail, a similar correlation has been found between the far-infrared emission and molecular column density inferred from CO (de Vries et al. 1987, Heiles, Reach, & Koo 1988; Reach, Koo, & Heiles 1994). These correlations suggest that most of the dust responsible for the far-infrared emission at high latitudes is heated fairly uniformly in both atomic and molecular clouds by stellar radiation from the Galactic disk. Here we will use far-infrared emission as a gas tracer in order to assess how much molecular gas could have been missed by our CO survey—owing perhaps to restricted sampling or a failure of CO to trace $H_2$.

To calibrate far-infrared emission as a gas tracer, we will assume that the total gas column density in regions free of CO emission is simply the H I column density derived from existing 21 cm surveys. An infrared map so calibrated will be differenced with a map of H I column density to yield a prediction of molecular column density. This approach is similar to that followed by Desert, Bazel, & Boulanger (1988) to search for molecular regions at high Galactic latitude, but as discussed below, the 21 cm, CO, and far-infrared surveys used in the present analysis are all of significantly higher quality than those available to Desert et al. Their analysis was recently updated and extended by Reach, Wall, & Odegard (1998) using the DIRBE far-infrared bands; their analysis, however, was confined to $|b| > 20°$, and thus overlaps the present analysis only in the latitude range $|b| = 20°–32°$.

The map of H I column density ($N_{HI}$) shown in Figure 7$a$ was derived primarily from the Leiden-Dwingeloo 21 cm survey (Hartmann & Burton 1997). In addition to significant improvements in spatial and velocity coverage, angular and velocity resolution, and sensitivity over previous large-scale surveys, this survey was carefully corrected for the effects of stray radiation, a correction of particular importance for the present analysis devoted to the region within 32° of the Galactic plane. The Leiden-Dwingeloo survey was supplemented below its declination limit of –30° with data from the Parkes 18 m telescope (Cleary, Heiles, & Haslam 1979, Kerr et al. 1986). Velocity-integrated intensities from the Dwingeloo and Parkes surveys were compared in the region of overlap and found to agree generally to 3% or better. Velocity-integrated intensities were converted to atomic column densities on the assumption that the 21 cm line is optically thin. Optical depth corrections are generally small except within ~2° of the plane toward the inner Galaxy, where, as discussed below, our $H_2$ prediction is somewhat unreliable in any case owing to large variations in dust temperature along the line of sight.

The map of 100 μm emission ($I_{100}$) shown in Figure 7$b$ is derived from a reprocessed composite of the DIRBE and IRAS all-sky maps by Schlegel, Finkbeiner, & Davis (1998), who combined those two surveys to exploit the superior calibration of the DIRBE survey and the superior resolution of the IRAS survey. They also removed zodiacal foreground emission and confirmed point sources, supplemented small regions missing from the IRAS map with DIRBE data, and, most significantly for the present analysis, they corrected the 100 μm flux at each point to that expected from dust at an arbitrary fixed temperature of 18.2 K. The dust effective temperature used in this color correction was derived from the ratio of DIRBE 100 μm to 240 μm flux, assuming an $\alpha = 2$ emissivity model for the dust. The map in Figure 7$b$ has also been smoothed to an angular resolution of 36' to match that of the 21 cm map in Figure 7$a$.

Corrected to a constant dust temperature and with point sources removed, the $I_{100}$ map in Figure 7$b$ should be proportional to total gas column density ($N_{tot}$), provided the dust temperature, the dust emissivity law, and the gas-to-dust ratio are all constant along the line of sight. The map of $N_{HI}/I_{100}$ in Figure 7$c$ suggests that these conditions roughly hold for most lines of sight off the Galactic plane: with regions of detected CO excluded (*white*), $N_{HI}/I_{100}$ is fairly constant at $0.9 \pm 0.4$ x $10^{20}$ cm$^{-2}$ MJy$^{-1}$ sr, with a 1-σ spread of only 50%. Localized minima in this ratio are indicative of unobserved molecular gas, while very large-scale variations are probably due to Galactic-scale gradients in dust temperature along the line of sight—generally positive in sign toward the inner Galaxy and negative toward the outer Galaxy. Such gradients are not accounted for by the simple dust temperature correction applied by Schlegel et al. (1998).

To account for these large-scale variations in the infrared-to-gas ratio, we will not adopt a constant $I_{100}/N_{tot} = <I_{100}/N_{HI}>$, but rather allow the ratio to vary on an angular scale that is large compared to that of molecular clouds. Specifically, we adopt for $I_{100}/N_{tot}$ at each point the Gaussian-weighted average $I_{100}/N_{HI}$ of all surrounding points without detected CO, the Gaussian having a width of 10° (FWHM). The molecular column density in the central regions of a fairly uniform, undetected molecular cloud ≳ 10° in size would be significantly underestimated by our procedure, but the edges of such a cloud would still be readily apparent in our predicted molecular map, and as shown below, we find no evidence for such large, undetected clouds. Our exclusion of all points with detected CO from the determination of the local mean $I_{100}/N_{tot}$—a refinement not applied by the previous, somewhat similar analyses of Desert et al. 1988 and Reach et al. 1998—allows us to predict fairly well molecular column densities even quite close to the Galactic plane.

In summary, to produce a map of predicted molecular column density, the infrared map (Fig. 7$b$) was multiplied by a heavily-smoothed version of the $I_{100}/N_{HI}$ map (Fig. 7c) to obtain a map of total gas column density, from which the H I column density (Fig. 7a) was subtracted. The result is shown in Figure 8b, scaled to the observed units of velocity-integrated CO intensity using $X \equiv N_{H_2}/W_{CO} = 1.8$ x $10^{20}$ cm$^{-1}$ K$^{-1}$ km$^{-1}$ s (see § 5). For comparison, Figure 8$a$ is our observed CO map smoothed to the same angular resolution (36') and displayed with the same color palette. The longitude profiles in Figure 8$c$, which are integrated over all latitudes at $|b| > 5°$, and the latitude profiles in Figure 9a provide more quantitative comparisons of the observed and predicted maps.

The quantitative agreement between the molecular clouds we observe in CO (Fig. 8*a*) and those predicted from the complete, unbiased infrared and 21 cm surveys (Fig. 8*b*) demonstrates that our composite CO survey is nearly complete, in spite of its restricted sampling.  All of the major local clouds—Taurus, the Polaris Flare, the Aquila Rift, Ophiuchus, R Corona Australis, Lupus, Chamaeleon, Orion, and the λ Orionis Ring—as well as most smaller features out of the plane, appear in both maps with similar intensities.  Over the entire area covered by the high-sensitivity Leiden-Dwingeloo 21 cm survey, few clouds appear in one map and not in the other. The agreement is not as detailed outside the Leiden-Dwingeloo coverage (below the dotted line in Fig. 8*b*); owing to the lower quality 21 cm data used there, the predicted map is noisier and apparently contains some spurious features.

Although the agreement between the observed and predicted maps is good on a large scale, the point-to-point dispersion between the two is significantly larger than can be accounted for by instrumental noise in the far-infrared, H I, or CO surveys.  Excluding the Galactic plane ($|b| < 5°$), where the prediction is expected to break down owing to dust temperature variations along the line of sight, the dispersion is ~50% of the observed intensity (Fig. 10), or ~1 K km s$^{-1}$ for typical emission features out of the plane. The noise in the CO map is at least several times lower (§ 2.1), and the contribution from instrumental noise in the IRAS and Leiden-Dwingeloo surveys is negligible.  The high dispersion is plausibly the result of variations in $I_{100}/N_{tot}$ on scales less than 10°, caused in turn by variations in the gas-to-dust ratio, and by dust temperature variations not accounted for by the simple infrared color correction applied by Schlegel et al.  Small-scale variations in the X factor may also contribute to the dispersion.

With an effective noise out of the Galactic plane of ~1 K km s$^{-1}$ at 36' resolution, the predicted map should be able to reveal at the level of 3-σ all high-latitude molecular clouds with an integrated brightness greater than 1 K km s$^{-1}$ deg$^2$ (corresponding to an H$_2$ mass of 9 M$_\odot$ for a typical high-latitude cloud distance of 100 pc).  About a dozen such predicted clouds lying outside the boundaries of our large-scale surveys have been mapped with the northern telescope over the past few years (§ 3.1), and nearly all show CO emission in good agreement with that predicted.  Below the plane near *l*=250° and 270° the predicted emission is suspect, both because it lies outside the Leiden-Dwingeloo 21 cm survey and because it lies toward Hα peaks in the Gum Nebula, a nearby H II region spanning an extraordinary 36° on the sky (Chanot & Sivan 1983).  More significant is the prediction of additional molecular gas around the Chamaeleon clouds.  As mentioned in § 3.1, we may well have missed a significant amount of molecular gas there, since the current sampling with the southern telescope is rather tightly confined to the three main clouds.

A simple estimate of the total amount of emission lying outside our current sampling can be obtained from Figure 9*b*, which compares the predicted latitude profile integrated over *observed* longitudes (as in Fig. 9*a*) with the same profile integrated over *all* longitudes; the difference (*shaded*) is the predicted emission lying outside our current sampling. Such emission amounts to only 2% of the total observed, or 15% of the emission observed at $|b| > 5°$; as just mentioned, it is mainly confined to the regions of Chamaeleon and the Gum Nebula.

## 5. The Local Molecular Gas Layer

The ratio of the predicted molecular column density (Fig. 8*b*) to the observed CO intensity (Fig. 8*a*) provides a calibration of the CO-to-$H_2$ mass conversion factor X. Since there are significant point-to-point variations in the derived X value, and because X might be expected to vary with latitude, we determine here only a mean X at each latitude by taking the ratio of mean $W_{CO}$ to mean predicted $N_{H_2}$ over the same observed positions—essentially the ratio of the curves in Figure 9*a*, but with the predicted curve in the derived units of $N_{H_2}$. The X value as a function of latitude is shown in Figure 11. The high X near the Galactic plane is probably spurious since, due to a lack of CO-free regions toward the inner plane, we cannot properly determine $I_{100}/N_{tot}$ there. Out of the Galactic plane ($|b| > 5°$), X shows little systematic variation with latitude from a mean value of $1.8 \pm 0.3 \times 10^{20}$ cm$^{-2}$ K$^{-1}$ km$^{-1}$ s.

Given the large sky area and large quantity of CO data involved, our calibration of X is perhaps the most reliable to date in the solar neighborhood. It agrees well with Galaxy-wide average values derived from analyses of the Galactic diffuse gamma ray emission ($1.9 \times 10^{20}$, Strong & Mattox 1996; $1.6 \times 10^{20}$, Hunter et al. 1997), and with a similar color-corrected infrared analysis of the Ursa Major and Camelopardalis clouds by Reach et al. (1998; $1.3 \pm 0.2 \times 10^{20}$). As Reach et al. (1998) suggest, the significantly lower X value derived for the Ursa Major clouds by de Vries et al. (1987), based on 100 μm data alone ($0.5 \pm 0.3 \times 10^{20}$), could be due to systematically lower dust temperatures within the molecular gas than in the surrounding atomic gas.

Assuming that our inventory of molecular gas at intermediate latitudes is nearly complete, it is of interest to compare our molecular column densities with those observed in the plane near the Sun and at higher latitudes. Figure 12 compares mean latitude profiles from both the CO observations and the infrared prediction (over all longitudes) with the profile expected for a plane parallel layer with a Gaussian thickness of 87 pc (FWHM) and a molecular surface density of 0.87 M$_\odot$ pc$^{-2}$, the values derived by Dame et al. (1987) from a study of molecular clouds within 1 kpc of the Sun (surface density scaled to the X value derived here). The observed steep column density decrease with latitude relative to that of a plane parallel layer is readily explained by the confinement of most of the local $H_2$ to relatively few large clouds, especially those in Orion, Taurus, and Ophiuchus, none of which is at high latitude. The sharp decline in molecular column density apparently continues to higher latitudes: the mean $N_{H_2}$ at $b >$ 30° implied by the northern Galactic hemisphere CO survey of Hartmann, Magnani, & Thaddeus (1998)[1], ~0.02 x $10^{20}$ cm$^{-2}$ , and the mean $N_{H_2}$ at $|b| > 20°$ implied by the infrared-excess analysis of Reach et al. (1998)[2], ~0.03 x $10^{20}$ cm$^{-2}$, are both consistent with an extrapolation of the infrared predicted curve in Figure 12.

## 6. Summary

Large-scale CO surveys of the Galactic plane and all large local clouds at higher latitudes have been combined into a composite survey of the entire Milky Way at an angular resolution ranging from 9' to 18'. The composite survey has up to 3.4 times higher angular resolution and up to 10 times higher sensitivity per unit solid angle than that of the only previous complete CO Galactic survey (Dame et al. 1987).

Panoramic spatial and longitude-velocity maps produced from the composite survey resolve the molecular Galaxy into hundreds of fairly well-defined giant molecular clouds that are mainly confined to the inner spiral arms of the Galaxy. Owing to less velocity crowding and a lower cloud density overall, the most obvious molecular spiral arms are those beyond and just inside the solar circle: specifically the Perseus arm in the second quadrant and the far side of the Carina arm in the fourth quadrant. Except for the prominent Cepheus and Polaris Flares, much of the local CO emission ($|v| <$ 20 km s$^{-1}$) appears to follow Gould's Belt, the apparent disk of OB stars, gas, and dust surrounding the Sun and inclined ~20° to the Galactic plane. The maximum northern extension of the Belt is marked by the clouds in Lupus, Ophiuchus, and Aquila while its southern extension is marked by those in Orion and Taurus.

The IRAS far-infrared survey of the Galaxy was used as a tracer of total gas column density, and calibrated as such with the Leiden-Dwingeloo 21 cm survey in regions free of CO emission. Once calibrated, the total-gas map was differenced with a map of H I column density derived from the 21 cm survey to obtain a complete and unbiased predicted map of molecular column density. The close agreement between this map and the observed $W_{CO}$ map implies that few molecular clouds at $|b| <$ 30° have been missed by the present CO survey, in spite of the gaps in its coverage away from Population I objects. In addition, the ratio of the observed CO map to the derived molecular map provided a measure of the local CO-to-$H_2$ mass conversion factor X. The mean value of

---

[1] Derived from Table 2 of Hartmann, Magnani, & Thaddeus (1998), assuming a survey completeness of 0.5 and X=1.8 x $10^{20}$. Hartmann et al. estimate the survey completeness to be in the range 0.3–0.7.

[2] Derived from Figure 6 of Reach, Wall, & Odegrad (1998), with assistance from W. Reach.

X at $|b| > 5°$ is $1.8 \pm 0.3 \times 10^{20}$ cm$^{-2}$ K$^{-1}$ km$^{-1}$ s, similar to the mean value derived for the Galaxy at large.

We thank L. Bronfman, S. W. Digel, J. May, S. T. Megeath, and B. A. Wilson for providing survey data in advance of publication, E. S. Palmer for his work in maintaining and upgrading the northern telescope over many years, H. Alvarez for his technical efforts on behalf of the southern telescope, and J. Dalton, T. G. Lippegrenfell, and A. J. Philbrick for substantial assistance with the data taking. We are indebted to A. Kerr and S.-K. Pan for their role in developing the sensitive and very dependable receivers of the 1.2 m telescopes. We also thank A. Mellinger for providing his optical panorama of the Milky Way in digital form.

Observations with the northern telescope since 1989 have been supported by Scholarly Studies grants from the Smithsonian Institution.

## A. Beam Efficiency and Size

$T_A{}^*$, the intensity given by chopper wheel calibration (Kutner & Ulich 1981), and $T_{mb}$, the main beam brightness temperature (Downes 1989), are related by $T_{mb} = T_A{}^* / \eta$, where $\eta$ is the main beam efficiency, the fraction of the forward power in the main beam.

Radiation issuing from the feed horn is either focused into the main beam ~5' in radius, diffracted by the primary, the secondary, and the support arms mainly into a region ~1° in radius, or misses the secondary mirror entirely and spills mainly into an annulus which extends from the edge of the secondary at an angle of 7.5° to ~15°. This spillover is readily calculated from the geometry of the telescope and the beam pattern of the feed horn. Using the computer code of Potter (1972), we calculated that 8% of the power is projected beyond the edge of the secondary. An additional 1% of the power is estimated to be lost in a small alignment hole in the secondary, so 91% of the total forward power illuminates the secondary.

The power incident on the secondary is assumed to illuminate the primary, to be diffracted by its finite aperture, and then by the secondary and the support arms. The diffraction pattern of the northern telescope, measured in late 1994 with a remote transmitter at $2.4\ D^2/\lambda$, is shown in Figure 13. Because the transmitter power was increased with offset from the main beam, this map is essentially noise-free even for sidelobes more than 50 dB below the main beam. The six-fold symmetry in the beam pattern is the expected diffraction from the three secondary support arms. More finely-spaced scans of the beam pattern in azimuth and elevation are compared to theoretical calculations based on scalar diffraction theory in Figures 14*a* and 14*c*. Our application of scalar diffraction theory followed Silver (1964)[3], as described by Cohen (1978)[4]. For the $(1-r^2)^p$ functional form of the aperture illumination function we used p=2, rather than p=1 as used by Cohen (1978), to better match the illumination of our current conical feed horn. By combining the measured beam pattern at offsets less than 3° with the theoretical beam pattern carried out to 20°, we estimated that 90% of the power illuminating the secondary is focused into the main beam.

To summarize, 91% of the forward power illuminates the secondary, and 90% of that is focused into the main beam, for a main beam efficiency of 0.82. The same value was estimated for the southern telescope by Bronfman et al. (1988). Gaussian fits to the azimuth and elevation scans plotted linearly (Figs. 14*b* and 14*d*) yield an average beamwidth of 8'.4 ± 0'.2 (FWHM) for the northern telescope.

---

[3] There is a factor 2 missing in eq. 75a of Chapter 6 in Silver (1964), as can be verified by integrating eq. 75 for the case p=1

[4] As a result of the error in eq. 75 of Chapter 6 of Silver (1964), the equations for $g_o$ and $g_c$ on p. 87 of Cohen (1978) are both missing a factor of 2 in front of the term involving $J_2$.

## B. Reference Positions (OFFs)

The position-switching we adopted (§ 2.1) requires two emission-free reference positions (OFFs) for each observation, one located above the source in elevation and one below. OFFs were typically chosen a few degrees or more from the Galactic plane and spaced every few degrees in Galactic longitude. In choosing candidate OFFs, known dark clouds were avoided; in recent years the IRAS 100 μm map (Fig. 7*b*) and the map of predicted molecular clouds (Fig. 8*b*) have proven to be reliable guides to regions free of CO emission. OFFs were checked to be CO-free by frequency switching to a noise level that was typically twice as low as that of the survey with which it would be used.

The positions of our 410 standard OFFs with respect to the Galactic CO emission are shown in Figure 15. The 4 OFFs at ($l$, $b$) = (18°, 4°), (23°, -4°), (24°, 4°), and (33°, 5°) were intentionally chosen within the large, nearby Aquila Rift cloud ($v \sim 10$ km s$^{-1}$), since no positions totally free of emission could be found sufficiently close to the plane in that region. Spectra obtained using one of these contaminated OFFs were corrected by adding in the Aquila Rift spectral line, weighted by the fraction of the time the OFF was used. Low-noise profiles for each of the contaminating lines were obtained by frequency switching. A complete list of the OFFs, including the *rms* level to which each was checked to be CO-free, is available on request.

**FIGURE CAPTIONS**

Fig. 1.— Areas covered by the individual CO surveys used to synthesize the composite survey of the Milky Way, numbered as in Table 1. A single contour at 2 K km s$^{-1}$ from a smoothed version of the velocity-integrated map (Fig. 8*a*) is overlaid.

Fig. 2.— Velocity-integrated CO map of the Milky Way. The angular resolution is 9' over most of the map, including the entire Galactic plane, but is lower (15' or 30') in some regions out of the plane (see Fig. 1 & Table 1). The sensitivity varies somewhat from region to region, since each component survey was integrated individually using moment masking or clipping in order to display all statistically significant emission but little noise (see § 2.2). A dotted line marks the sampling boundaries, given in more detail in Fig. 1. [NOTE: *This figure will be on the front side of a 5-page color foldout*]

Fig. 3.— Longitude-velocity map of CO emission integrated over a strip ~4° wide in latitude centered on the Galactic plane (see § 2.2)—a latitude range adequate to include essentially all emission beyond the Local spiral arm (i.e., at $|v| > 20$ km s$^{-1}$). The map has been smoothed in velocity to a resolution of 2 km s$^{-1}$ and in longitude to a resolution of 12'. The sensitivity varies somewhat over the map, since each component survey was integrated individually using moment masking at the 3-σ level (see § 2.2).
[NOTE: *This figure will be on the back side of a 5-page color foldout*]

Fig. 4.— Comparison of the CfA Nyquist-sampled CO survey of the second quadrant with the FCRAO 14 m CO survey (Heyer et al. 1998) in a large region of overlap between the two. (*a*) W$_{CO}$ map from the FCRAO 14 m survey; the area shown comprises about half the entire FCRAO survey. (*b*) W$_{CO}$ map from the CfA Nyquist-sampled CO survey (No. 17 in Table 1). (*c*) same FCRAO map as in (a), but spatially smoothed to 9' resolution and scaled by 0.83 to match the resolution and absolute calibration of the CfA survey in (*b*). All three maps are displayed with the same color palette.

Fig. 5.— Longitude-velocity contours for large, local clouds lying outside the latitude integration range of the in-plane longitude-velocity map (Fig. 3), reproduced here in grayscale. The contour map has been smoothed in velocity to a resolution of 1.5 km s$^{-1}$ and in longitude to a resolution of 12'. The full map is divided by quadrants: (*a*) first quadrant, (*b*) second quadrant, (*c*) third quadrant, and (*d*) fourth quadrant. Contours are spaced at 3.5 K deg, starting at 0.35 K deg.

Fig. 6.— (*a*) Half of the optical panorama of the Milky Way assembled by A. Mellinger (di Cicco 1999), covering the first and fourth Galactic quadrants. The panorama is a composite of 16 optical photographs projected onto a rectangular Galactic coordinate grid.

In this image, the most intense regions toward the Galactic bulge have been saturated (set to white in the grayscale here) to better display the dark clouds. (*b*) A spatial map of molecular clouds with near kinematic distances less than 2.5 kpc according to the rotation curve of Burton (1988), obtained by integrating over a longitude-dependent velocity window. The window always included the range $v = -10$ to $10$ km s$^{-1}$ to account for the random motions of local clouds, but in the central portions of the first and fourth quadrants was extended as high as ~35 km s$^{-1}$ in absolute value. The resulting map was spatially smoothed to an angular resolution of 0.25°; contours are at 2, 8, 16, and 50 K km s$^{-1}$. An earlier version of this figure based on CO data with lower sensitivity and angular resolution is shown in Figure 4 of Dame & Thaddeus (1985).

Fig. 7.— (*a*) Velocity-integrated 21 cm line intensity, mainly from the Leiden-Dwingeloo survey (Hartmann & Burton 1997), but supplemented in the southern sky (below the dotted line in *b*) with data from the Parkes 18 m telescope (Cleary et al. 1979; Kerr et al. 1986).
(*b*) IRAS 100 μm intensity with zodiacal emission and confirmed point sources removed; this map has been supplemented, recalibrated, and color corrected to a reference dust temperature of 18.2 K using the DIRBE 100 μm and 240 μm maps. Specifically, this is the $D^T$ map of Schlegel et al. (1998; see their eq. 22), smoothed to an angular resolution of 36' to match that of the H I map in (*a*).
(*c*) The ratio of gas column density to 100 μm intensity in regions without detected CO. This map is essentially the ratio of those in (*a*) and (*b*), but with 21 cm velocity-integrated intensity converted to H I column density under the usual assumption of low optical depth: $N_{HI}$ (cm$^{-2}$) = 1.822 x 10$^{18}$ $W_{HI}$ (K km s$^{-1}$). The CO velocity-integrated map (Fig. 2) was smoothed to an angular resolution of 36' to match that of the H I map in (*a*), and all positions in the smoothed map (Fig. 8*a*) with $W_{CO} > 1$ K km s$^{-1}$ were blanked (displayed white) in the ratio map.

Fig. 8.— A comparison of the observed $W_{CO}$ map with a complete and unbiased prediction that uses far-infrared emission as a total gas tracer and 21 cm emission to account for the atomic gas.
(*a*) The observed $W_{CO}$ map (Fig. 2) smoothed to an angular resolution of 36' to match that of the predicted map in (*b*).
(*b*) The predicted $W_{CO}$ map obtained by multiplying the far-infrared map in Fig. 7*b* by a heavily-smoothed (FWHM=10°) version of the $N_{HI}/I_{100}$ map in Fig. 7*c* to yield a map of total gas column density, then subtracting from that the H I column density derived from Fig. 7*a*. The resulting map of $N_{H_2}$ was scaled to $W_{CO}$ using $X \equiv N_{H_2}/W_{CO} = 1.8$ x 10$^{20}$ cm$^{-1}$ K$^{-1}$ km$^{-1}$ s (see § 5 and Fig. 11). The dotted curve at δ = −30° marks the edge of the high-quality Leiden-Dwingeloo 21 cm survey (Hartmann & Burton 1997); below this curve the predicted map is less reliable owing to higher noise and stripping in the Parkes 21 cm data.

(*c*) Longitude profiles at $|b| > 5°$ derived from the observed and predicted $W_{CO}$ maps in (*a*) and (*b*), respectively. The sharp predicted peak labeled LMC is mainly due to the enormous 30 Dor H II region in that irregular galaxy; CO emission from the LMC (Cohen et al. 1988) is not included in the observed map in (*a*).

Fig. 9.— (*a*) Latitude profiles derived from the observed and predicted $W_{CO}$ maps in Figs. 8*a* and 8*b*, respectively. Both curves are integrated over all longitudes, and both exclude unobserved regions.

(*b*) Latitude profiles derived from the predicted $W_{CO}$ map in Fig. 8*b*. In one case (*solid curve*) the longitude integral is over all positions, while in the other (*dotted*) the integral is only over positions observed in CO. The difference between the two curves (*shaded*) represents predicted emission in unobserved regions.

Fig. 10.— The dispersion between the observed and predicted $W_{CO}$ maps (Figs. 8*a* and 8*b*, respectively) as a fraction of the observed $W_{CO}$. Only points with detected CO ($W_{CO} > 1$ K km s$^{-1}$) at $|b| > 5°$ are included. The Gaussian fit indicates that the predicted map has an uncertainty per half-degree pixel of about half the predicted intensity, or ~1 K km s$^{-1}$ for typical emission features out of the plane.

Fig. 11.— Mean $X \equiv N_{H_2}/W_{CO}$ as a function of Galactic latitude, derived by taking the ratio of predicted molecular column density (Fig. 8*b*) to observed CO intensity (Fig. 8*a*). This curve is simply the ratio of the latitude profiles in Fig. 9*a*, but with the predicted curve in its derived units of $N_{H_2}$ (cm$^{-2}$). The error bars assume an average noise per half-degree pixel of 0.34 K km s$^{-1}$ for the observed $W_{CO}$ map (*measured*) and the larger of 1.8 x 10$^{20}$ cm$^{-2}$ or $N_{H_2}/2$ for the predicted $N_{H_2}$ map (see § 4 and Fig. 10).

Fig. 12.— Mean molecular column density vs. $|b|$ derived from the observed and predicted $W_{CO}$ maps in Figs. 8*a* and 8*b*, respectively, using X = 1.8 x 10$^{20}$ cm$^{-2}$ K$^{-1}$ km$^{-1}$ s (see § 5). The predicted curve is averaged over all longitudes, even those not observed in CO. The error bars on the predicted curve assume an *rms* error per half-degree pixel of half the predicted $N_{H_2}$, or 1.8 x 10$^{20}$ cm$^{-2}$, whichever is larger. The dotted curve is the profile expected for a plane parallel layer with a Gaussian thickness of 87 pc (FWHM) and a molecular surface density of 0.87 M$_\odot$ pc$^{-2}$, the values derived by Dame et al. (1987) from an inventory of molecular clouds within 1 kpc of the Sun (the Dame et al. surface density is scaled to the $N_{H_2}/W_{CO}$ value used here).

Fig. 13.— Beam pattern of the northern telescope, measured in 1994 with a remote transmitter (Dame 1995). The main beam (FWHM) is marked by a white circle. The radial spokes are caused by diffraction from the three support arms of the secondary.

Fig. 14.— Azimuth and elevation scans of the beam pattern of the northern telescope, measured in 1994 with a remote transmitter (Dame 1995). (*a*) and (*c*) compare the azimuth and elevations scans plotted logarithmically (solid curves) with calculations based on scalar diffraction theory (dotted; Cohen 1978 & Dame 1995, based on the theory outlined in Silver 1964). (*b*) and (*d*) show the same scans plotted linearly (solid curves) and fitted with Gaussians (dotted).

Fig. 15.— Positions of the 410 emission-free reference positions (OFFs) used for the composite survey. Regions with $W_{CO} > 2.5$ K km s$^{-1}$ in the smoothed spatial map (Fig. 8*a*) are shaded gray.

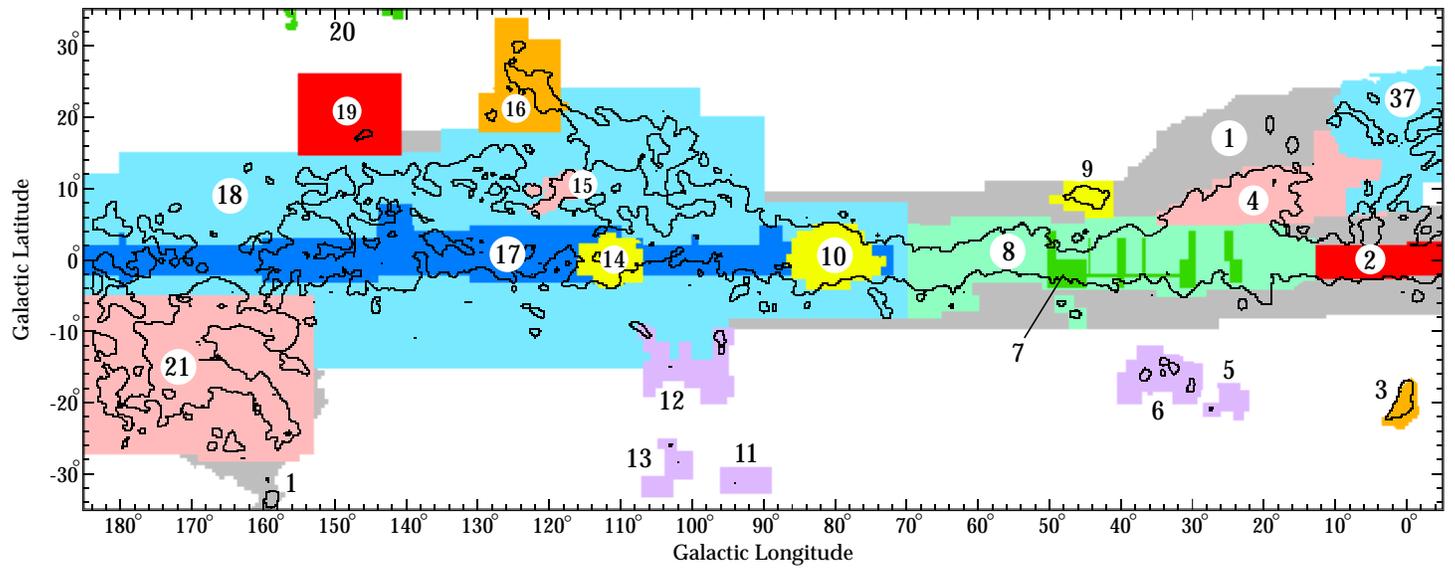

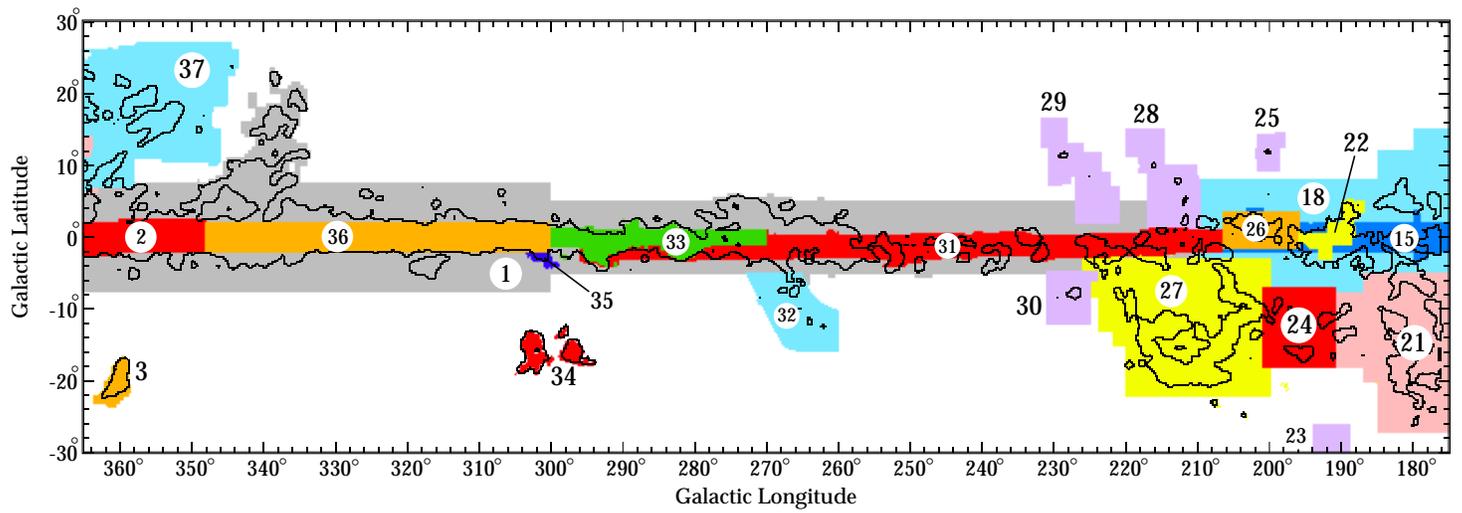

Figure 1

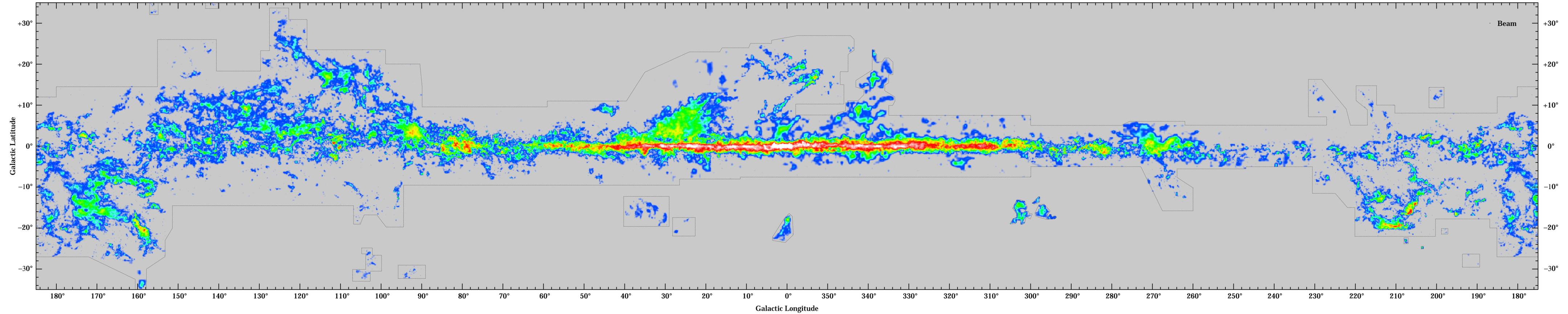

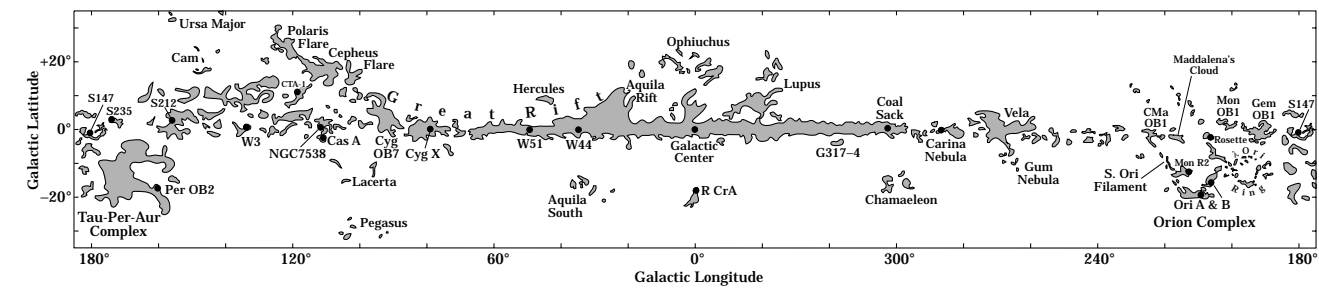

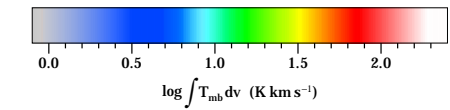

Fig. 2.—Velocity-integrated CO map of the Milky Way. The angular resolution is 9′ over most of the map, including the entire Galactic plane, but is lower (15′ or 30′) in some regions out of the plane (see Fig. 1 & Table 1). The sensitivity varies somewhat from region to region, since each component survey was integrated individually using moment masking or clipping in order to display all statistically significant emission but little noise (see §2.2). A dotted line marks the sampling boundaries, given in more detail in Fig. 1.

$\log \int T_{mb}\,dv$   (K km s$^{-1}$)

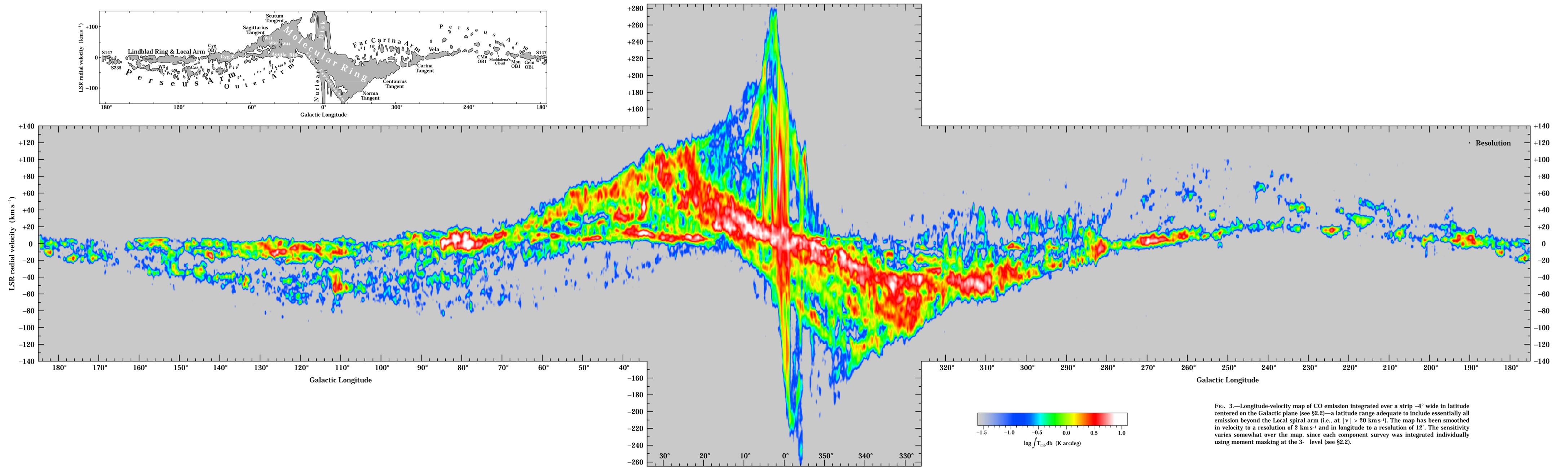

Fig. 3.—Longitude-velocity map of CO emission integrated over a strip ~4° wide in latitude centered on the Galactic plane (see §2.2)—a latitude range adequate to include essentially all emission beyond the Local spiral arm (i.e., at |v| > 20 km s⁻¹. The map has been smoothed in velocity to a resolution of 2 km s⁻¹ and in longitude to a resolution of 12′. The sensitivity varies somewhat over the map, since each component survey was integrated individually using moment masking at the 3-σ level (see §2.2).

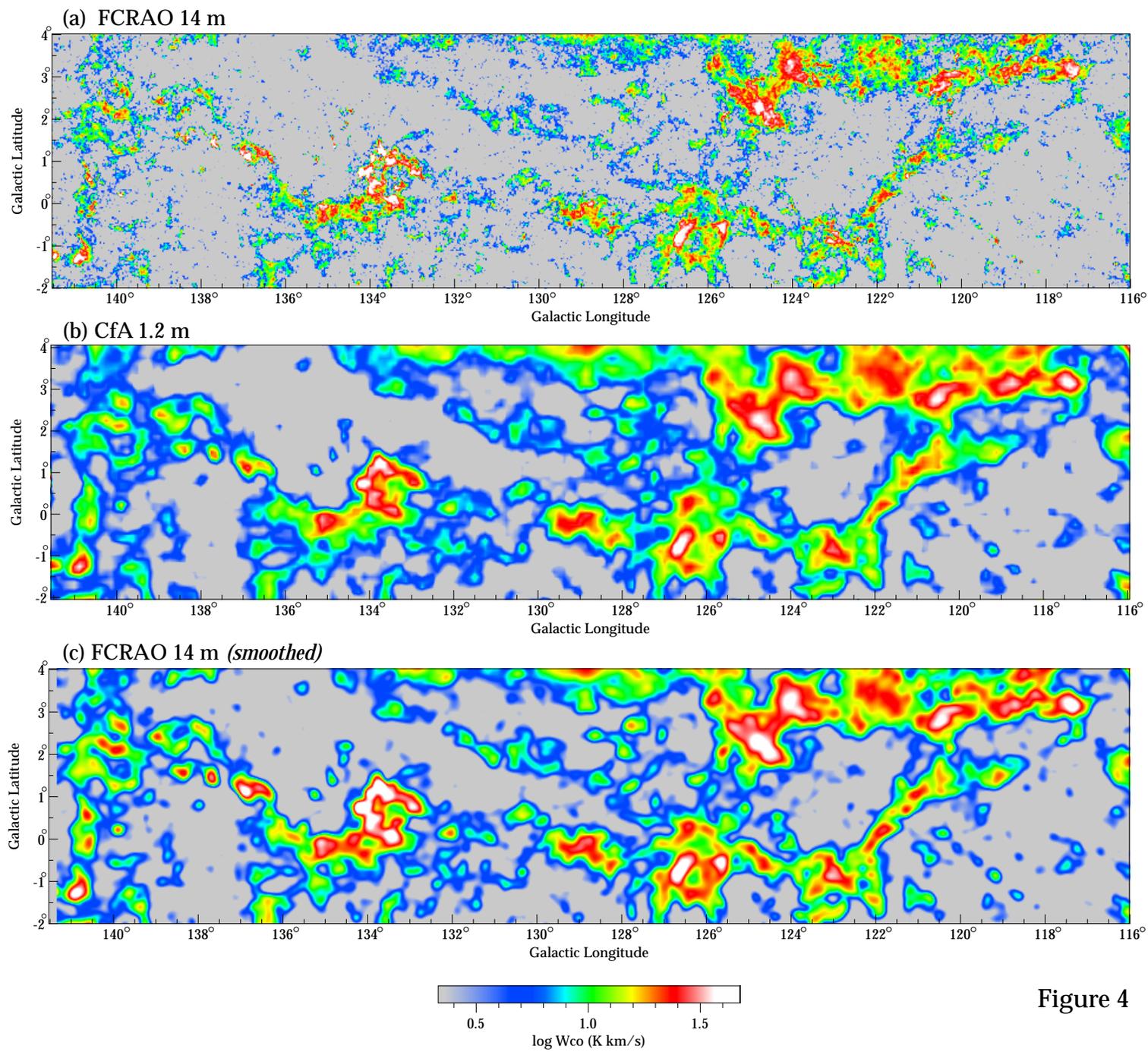

(a) FCRAO 14 m

(b) CfA 1.2 m

(c) FCRAO 14 m *(smoothed)*

log Wco (K km/s)

Figure 4

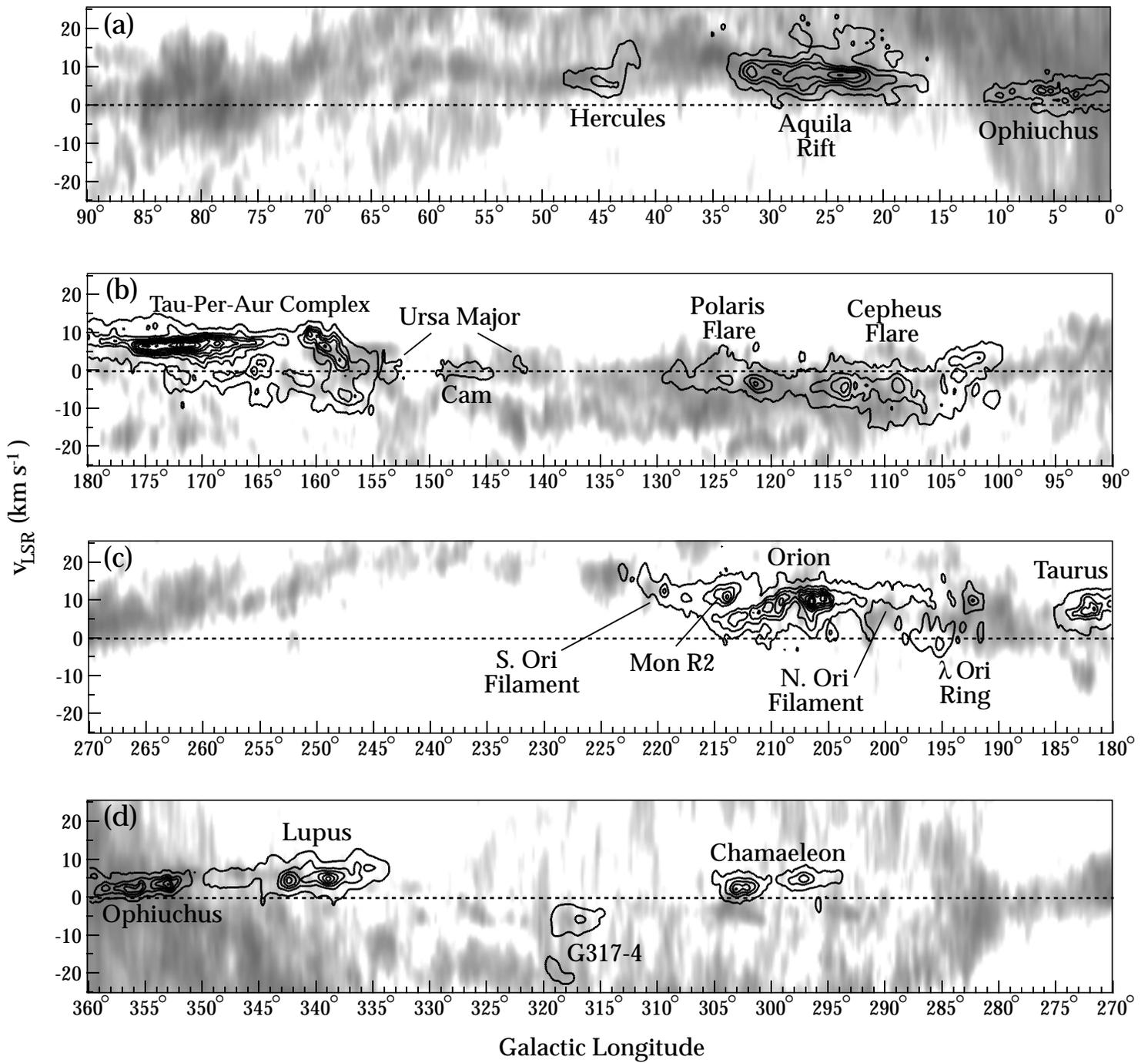

Figure 5

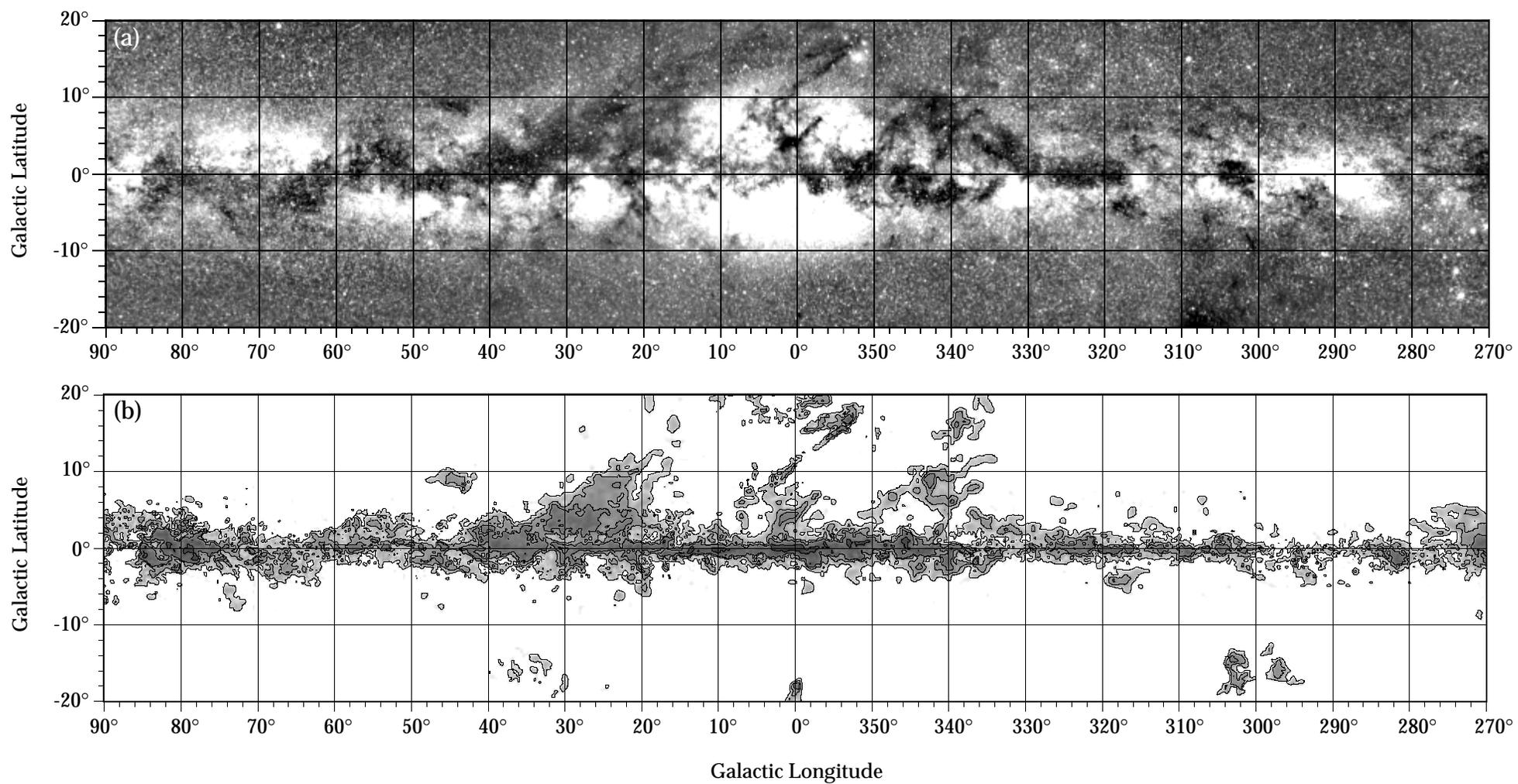

Figure 6

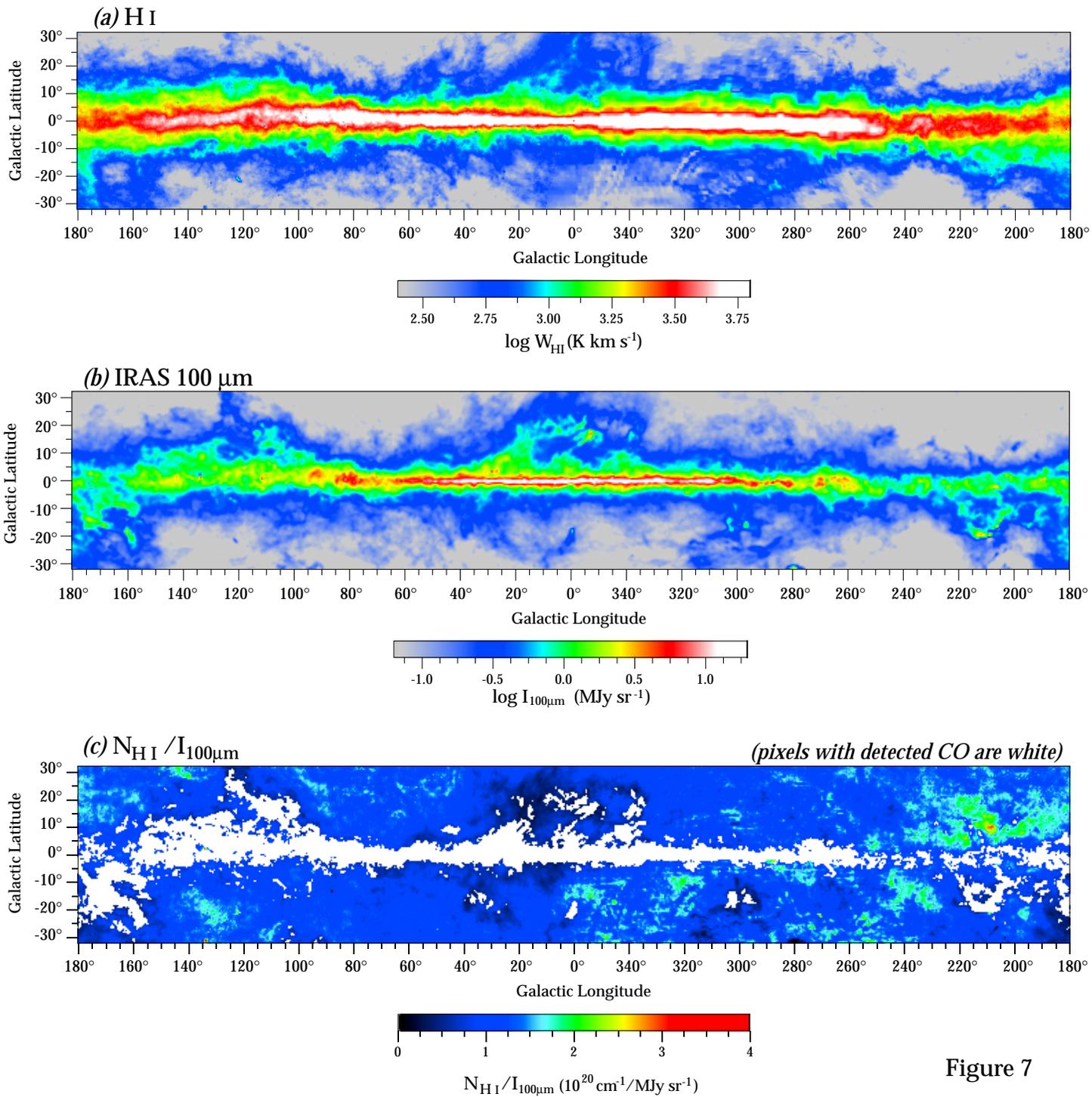

**(a)** H I

**(b)** IRAS 100 μm

**(c)** $N_{HI}/I_{100\mu m}$ *(pixels with detected CO are white)*

$\log W_{HI}$ (K km s$^{-1}$)

$\log I_{100\mu m}$ (MJy sr$^{-1}$)

$N_{HI}/I_{100\mu m}$ ($10^{20}$ cm$^{-1}$/MJy sr$^{-1}$)

Figure 7

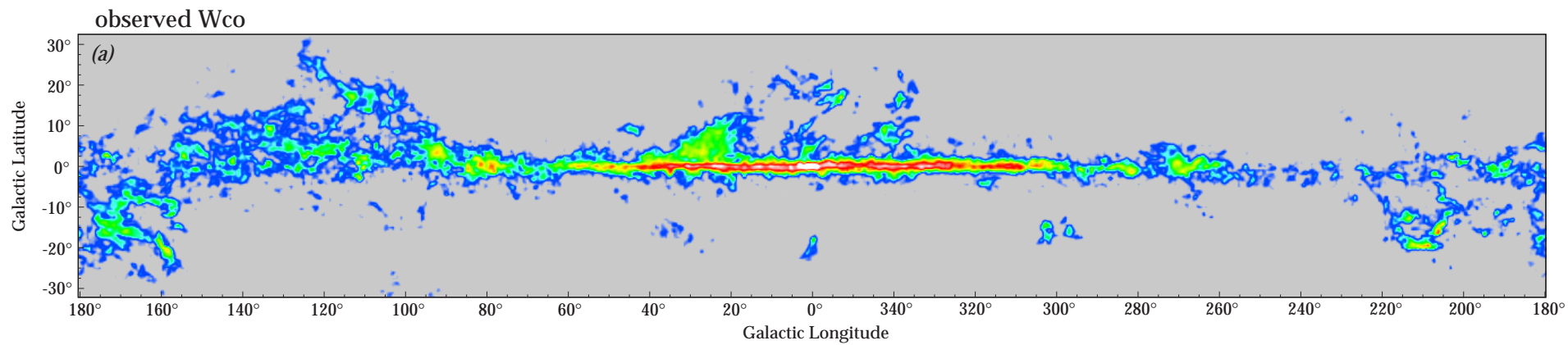

observed Wco

(a)

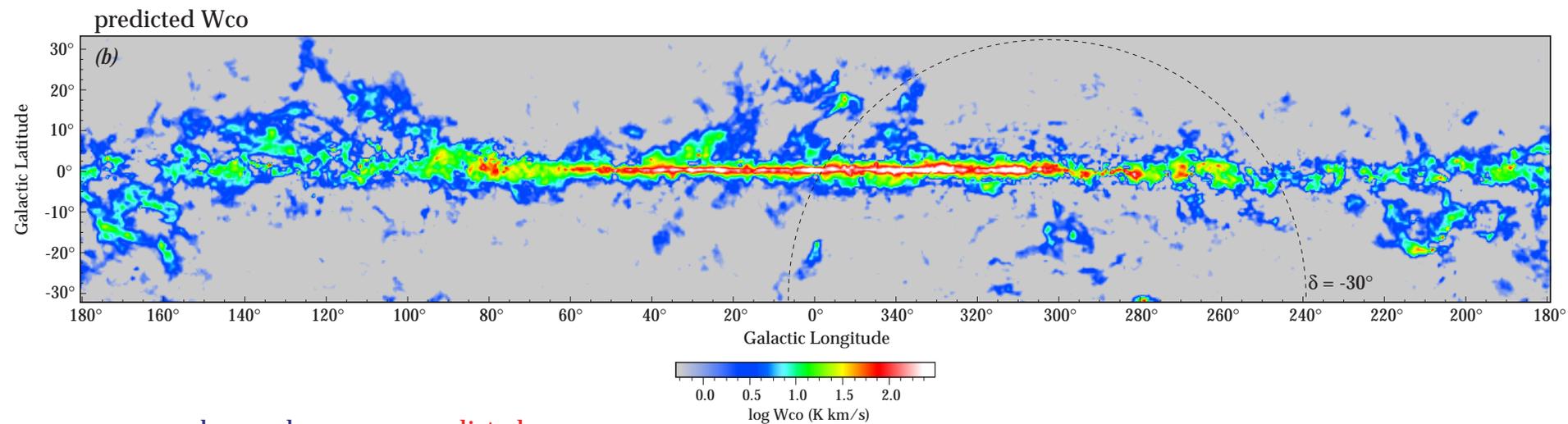

predicted Wco

(b)

δ = -30°

log Wco (K km/s)

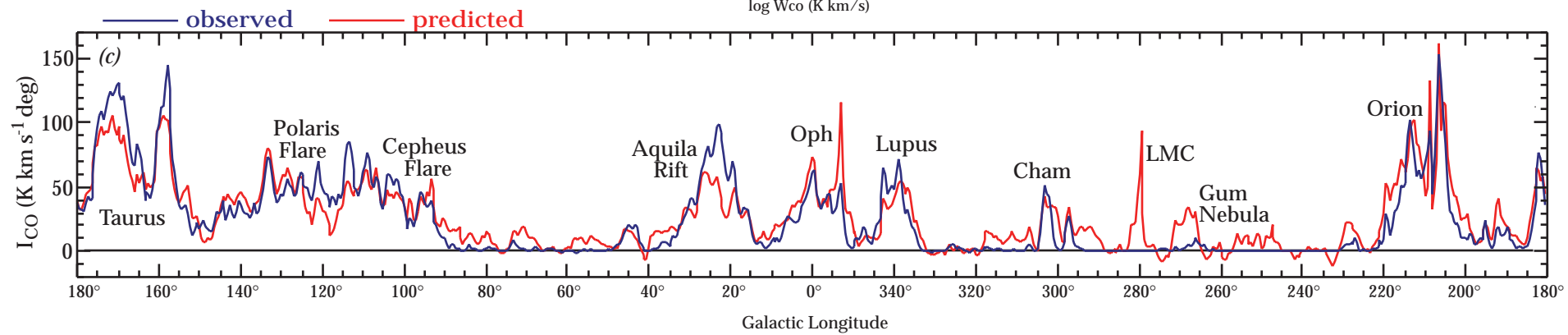

observed    predicted

(c)

Taurus    Polaris Flare    Cepheus Flare    Aquila Rift    Oph    Lupus    Cham    LMC    Gum Nebula    Orion

Figure 8

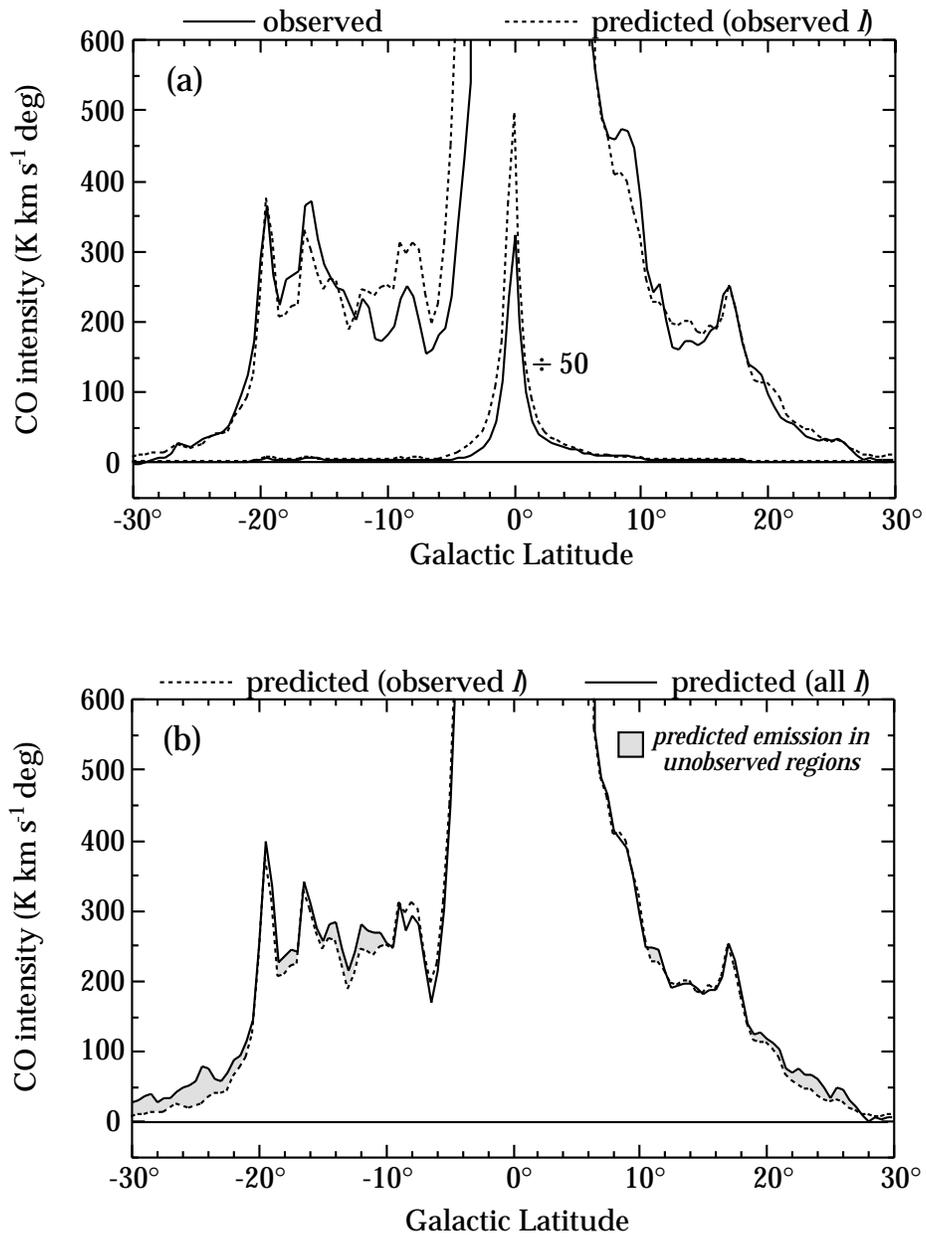

Figure 9

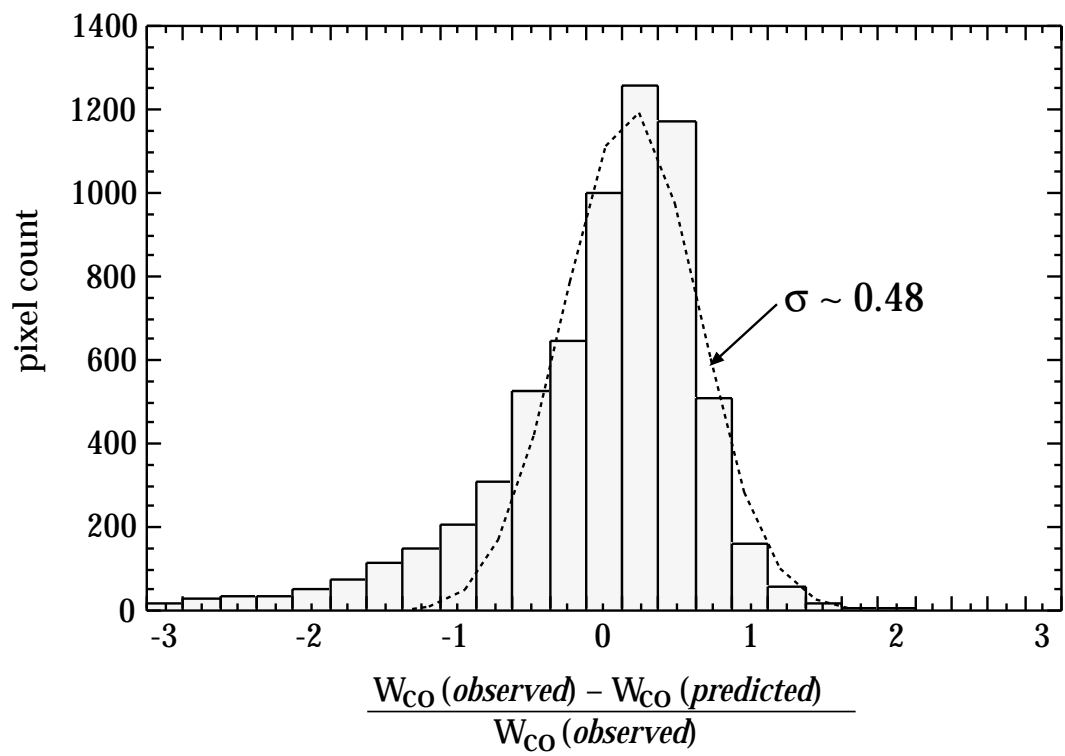

Figure 10

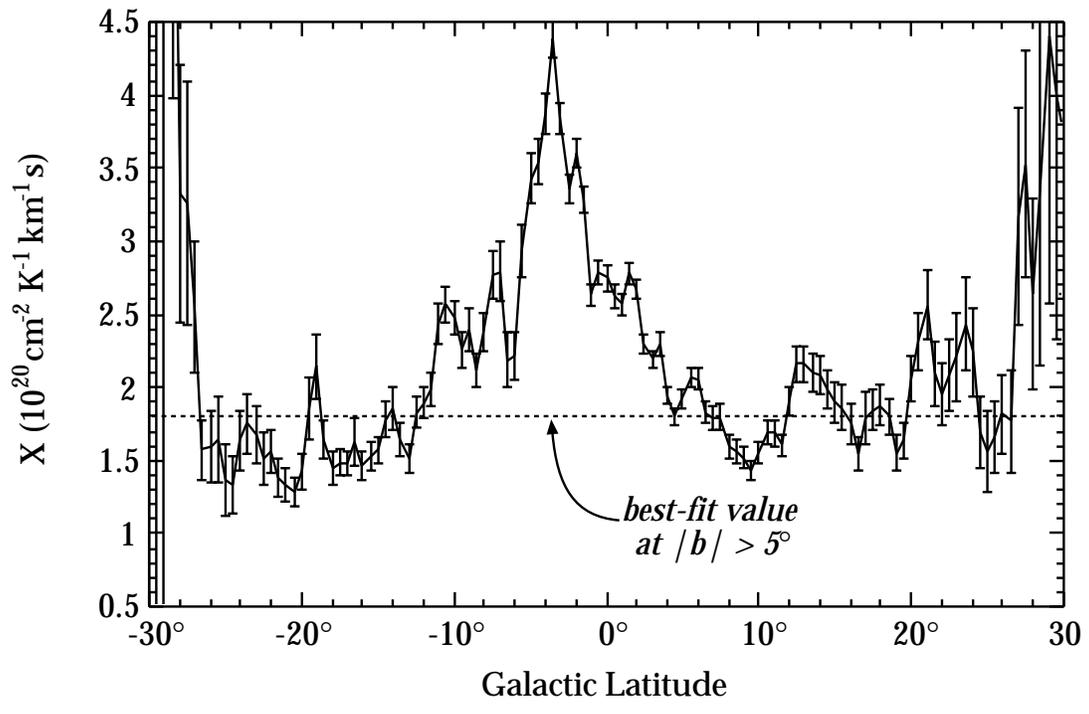

Figure 11

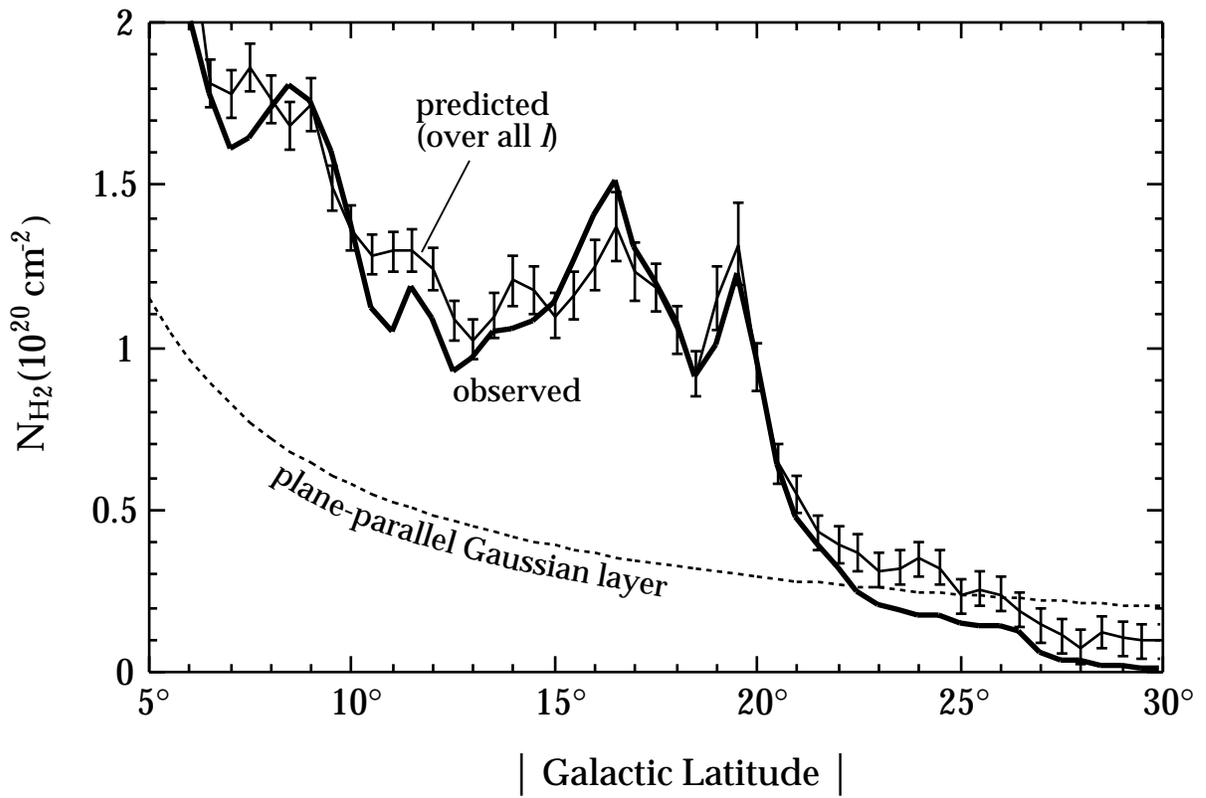

Figure 12

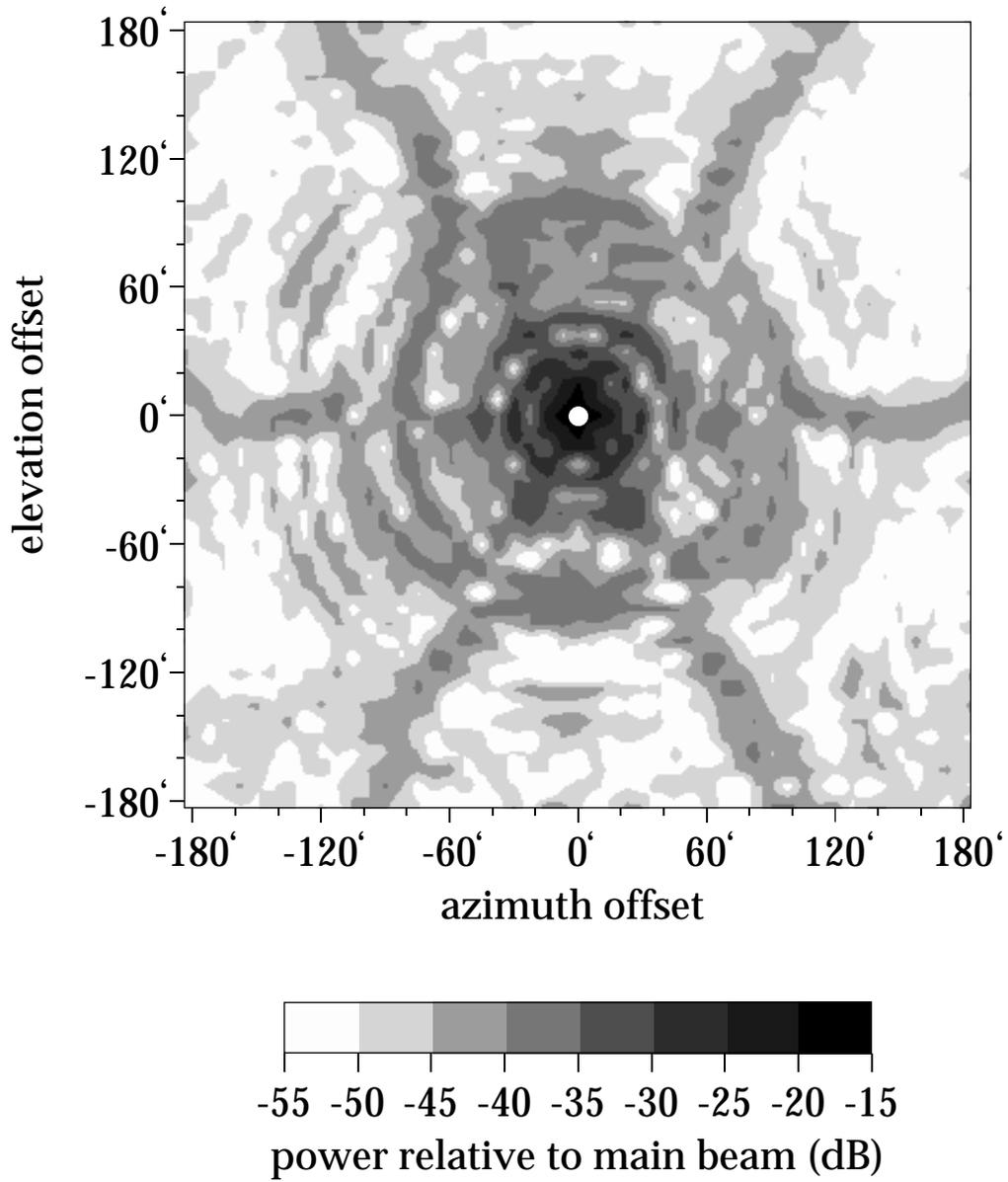

power relative to main beam (dB)

Figure 13

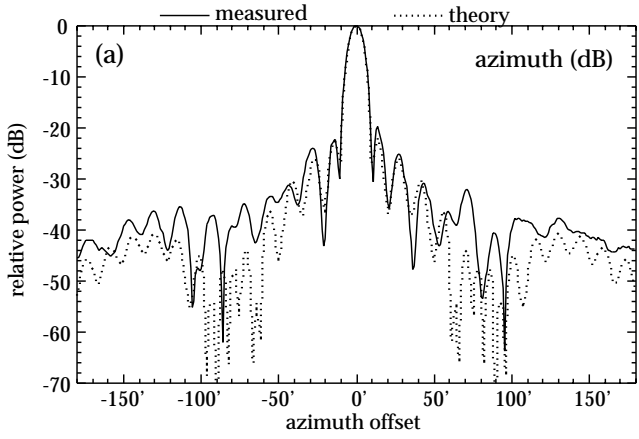

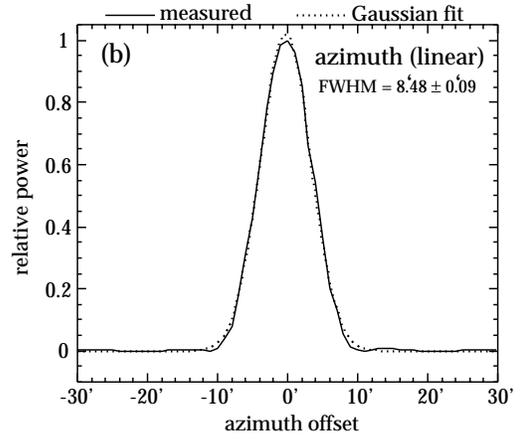

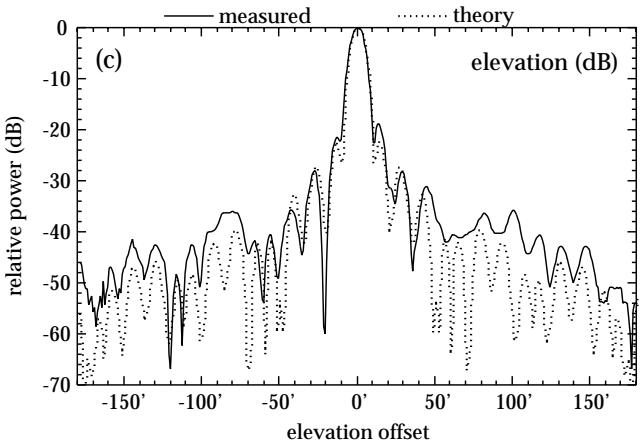

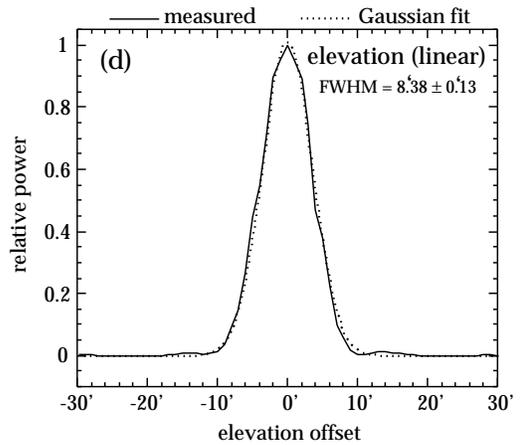

Figure 14

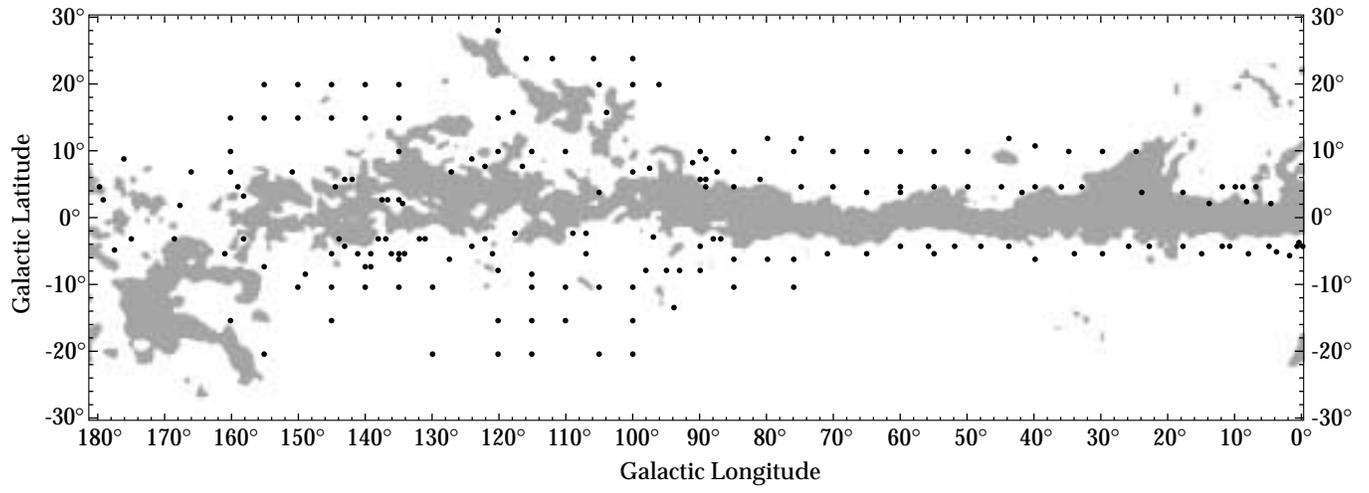

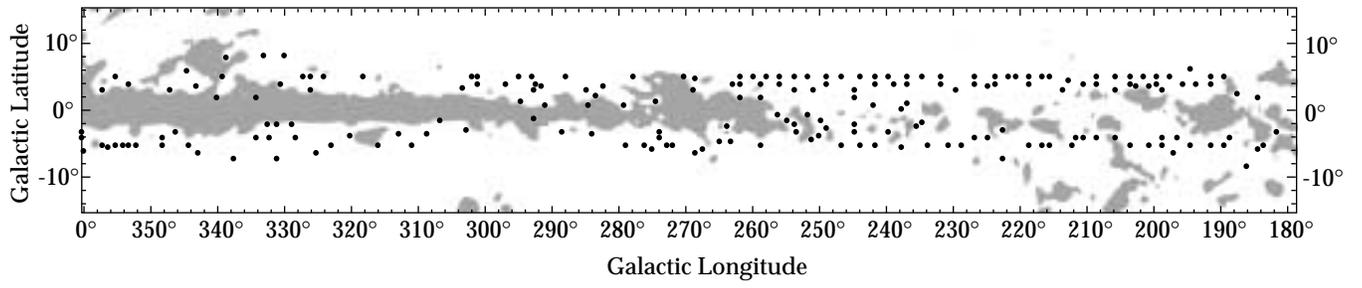

Figure 15

| No. | Region | $l$ (°) | $b$ (°) | Total Points | Area (deg$^2$) | $\Delta\theta^a$ (°) | $\Delta v^b$ (km s$^{-1}$) | $\sigma^c$ (K) | $\sigma'^d$ (K) | Tel. | Ref. |
|---|---|---|---|---|---|---|---|---|---|---|---|
| 1 | Superbeam | 0 | 0 | 28,107 | 7,027 | 0.5 | 0.65 | 0.25 | 1.00 | NY/Chile | 1 |
| 2 | Gal. Center | 0 | 0 | 4,492 | 117 | 0.125 | 1.30 | 0.12 | 0.17 | Chile | 2 |
| 3 | R CrA | 0 | -21 | 344 | 21 | 0.25 | 0.26 | 0.30 | 0.38 | Chile | 1 |
| 4 | Aquila Rift | 20 | 8 | 4,379 | 227 | 0.25 | 0.65 | 0.18 | 0.36 | CfA | 3 |
| 5 | Sag-South | 26 | -20 | 359 | 22 | 0.25 | 0.65 | 0.26 | 0.52 | CfA | 3 |
| 6 | Aql-South | 35 | -16 | 1,178 | 74 | 0.25 | 0.65 | 0.26 | 0.52 | CfA | 3 |
| 7 | 1st Quad | 37 | 0 | 15,916 | 62 | 0.0625 | 0.65 | 0.18 | 0.09 | CfA | 4 |
| 8 | 1st Quad | 45 | 0 | 37,610 | 595 | 0.125 | 0.65 | 0.18 | 0.18 | CfA | 3 |
| 9 | Hercules | 45 | 9 | 2,247 | 35 | 0.125 | 0.65 | 0.18 | 0.18 | CfA | 5 |
| 10 | Cygnus X | 81 | 0 | 6,029 | 94 | 0.125 | 0.65 | 0.12 | 0.12 | CfA | 6 |
| 11 | Peg-West | 93 | -31 | 1,653 | 26 | 0.125 | 0.65 | 0.26 | 0.26 | CfA | 3 |
| 12 | Lacerta | 101 | -16 | 5,355 | 84 | 0.125 | 0.65 | 0.26 | 0.26 | CfA | 3 |
| 13 | Peg-East | 103 | -29 | 535 | 33 | 0.25 | 0.65 | 0.15 | 0.30 | CfA | 3 |
| 14 | Cas A | 111 | 0 | 3,533 | 55 | 0.125 | 0.65 | 0.15 | 0.15 | CfA | 7 |
| 15 | SNR CTA 1 | 119 | 10 | 1,804 | 28 | 0.125 | 0.65 | 0.43 | 0.43 | NY | 8 |
| 16 | Polaris Flare | 123 | 24 | 5,291 | 134 | 0.125 | 0.65 | 0.13 | 0.13 | CfA | 9 |
| 17 | 2nd Quad | 135 | 0 | 146,944 | 574 | 0.0625 | 0.65 | 0.31 | 0.16 | CfA | 3 |
| 18 | 2nd Quad | 135 | 0 | 45,232 | 2,827 | 0.25 | 0.65 | 0.31 | 0.62 | CfA | 3 |
| 19 | Camelopardalis | 148 | 20 | 3,013 | 159 | 0.125 | 0.65 | 0.13 | 0.13 | CfA | 9 |
| 20 | Ursa Major | 148 | 35 | 2,118 | 44 | 0.125 | 0.65 | 0.13 | 0.13 | CfA | 9 |
| 21 | Taurus | 170 | -15 | 56,522 | 883 | 0.125 | 0.65 | 0.25 | 0.25 | CfA | 10 |
| 22 | Gem OB1 | 191 | 0 | 2,009 | 31 | 0.125 | 0.65 | 0.12 | 0.12 | CfA | 11 |
| 23 | Tau-South | 192 | -28 | 357 | 22 | 0.25 | 0.65 | 0.26 | 0.52 | CfA | 3 |
| 24 | Lam Ori | 196 | -13 | 28,885 | 113 | 0.0625 | 0.65 | 0.25 | 0.13 | CfA | 12 |
| 25 | Gemini | 200 | 12 | 1,165 | 18 | 0.125 | 0.65 | 0.26 | 0.26 | CfA | 3 |
| 26 | Mon OB1 | 201 | 1 | 13,440 | 52 | 0.0625 | 0.65 | 0.24 | 0.12 | CfA | 13 |
| 27 | Orion | 212 | -9 | 28,342 | 443 | 0.125 | 0.65 | 0.26 | 0.26 | CfA | 14 |
| 28 | Canis Minor | 214 | 6 | 6,922 | 108 | 0.125 | 0.65 | 0.26 | 0.26 | CfA | 3 |
| 29 | Monoceros | 227 | 10 | 1,442 | 90 | 0.25 | 0.65 | 0.26 | 0.52 | CfA | 3 |
| 30 | Canis Major | 228 | -8 | 774 | 48 | 0.25 | 0.65 | 0.26 | 0.52 | CfA | 3 |
| 31 | 3rd Quad | 245 | -1 | 6,750 | 356 | 0.25 | 1.30 | 0.12 | 0.34 | Chile | 15 |
| 32 | Gum Nebula | 266 | -10 | 1,553 | 97 | 0.25 | 0.26 | 0.25 | 0.32 | Chile | 16 |
| 33 | Carina | 285 | 0 | 6,162 | 96 | 0.125 | 1.30 | 0.17 | 0.24 | Chile | 17 |
| 34 | Chamaeleon | 300 | -16 | 1,744 | 27 | 0.125 | 1.30 | 0.30 | 0.42 | Chile | 18 |
| 35 | Coalsack | 303 | 0 | 1,736 | 27 | 0.125 | 0.26 | 0.30 | 0.19 | Chile | 19 |
| 36 | 4th Quad | 324 | 0 | 7,195 | 192 | 0.125 | 1.30 | 0.12 | 0.17 | Chile | 20 |
| 37 | Ophiuchus | 355 | 17 | 6,753 | 422 | 0.25 | 0.26 | 0.31 | 0.39 | Chile | 21 |

[a] sampling interval in longitude and latitude
[b] spectrometer channel width
[c] rms noise per channel
[d] rms noise per unit solid angle and velocity interval: $\sigma' = \sigma \, (\Delta\theta/\Delta\theta_0) \, (\Delta v/\Delta v_0)^{-1/2}$, where $\Delta\theta_0 = 0°.125$ and $\Delta v_0 = 0.65$ km s$^{-1}$